\documentclass[preprint,12pt,number,sort&compress]{elsarticle}
\usepackage{amsmath,amssymb,bm}
\usepackage{graphicx}
\usepackage{booktabs}
\usepackage{tikz}
\usetikzlibrary{arrows.meta,positioning,patterns}
\usepackage[colorlinks=true,allcolors=blue]{hyperref}
\AtBeginDocument{\DeclareTextSymbolDefault{\textbullet}{OMS}}   % keep the bullet in a Type 1 font
\journal{Journal of Sound and Vibration}
\usepackage{etoolbox}
\makeatletter\patchcmd{\pprintMaketitle}{\par\vskip36pt}{\par\vskip18pt}{}{\typeout{PATCH FAILED}}\makeatother

\newcommand{\rr}{\bm{r}}
\newcommand{\that}{\hat{\bm{t}}}
\newcommand{\eps}{\varepsilon}

\begin{document}

\begin{frontmatter}

\title{A comparison of three models of a whirling string as it becomes
chaotic}

\author{Nicholas B. Tufillaro\corref{cor1}}
\ead{nick@aqualytics.eco}
\cortext[cor1]{Corresponding author.}
\affiliation{organization={Aqualytics}, city={Corvallis}, state={Oregon}, country={USA}}

\begin{abstract}
A stretched wire driven near its fundamental frequency can leave the plane
of the driving force and whirl. At larger amplitudes the whirl becomes
modulated, reverses its direction, and becomes chaotic. Three models of this
motion are in common use: the geometrically exact string equations; a
two-mode model---the Kirchhoff--Carrier equation with one transverse mode in
each plane; and the averaged equations for the slow amplitudes of those two
modes. We integrate all three under similar conditions and compare them
along the route to chaos, which starts with a Hopf bifurcation of the steady
whirl and passes through period doublings to chaotic attractors organized by
a Shil'nikov homoclinic orbit. The small parameter of the averaging, the
ratio $\varepsilon = c/\omega_1$ of the modulation rate to the fundamental
frequency, runs from 0.005 to 0.02. Qualitatively the three models agree in
the sequence of bifurcations, the coexisting attractors, the return maps and
the Lyapunov exponents. Quantitatively, averaging places each local
bifurcation $0.7\varepsilon$ to $1.2\varepsilon$ higher in the scaled
detuning. The geometrically exact and two-mode models agree with each other
to within 0.002 at the local bifurcations for $\varepsilon \le 0.01$. The
Shil'nikov ratio, which decides whether chaos accompanies the homoclinic orbit,
agrees among the three models to within 1.5\% for $\varepsilon \le 0.01$,
yet the homoclinic bifurcation itself moves by about $\varepsilon$ at the
lower damping and by 6 to 9 times $\varepsilon$ at the higher; the
difference is set by how quickly the return of the homoclinic loop changes
with detuning.
\end{abstract}


\end{frontmatter}

\section{Introduction}
\label{sec:intro}

A wire stretched between two supports, and weakly driven near its
fundamental frequency, vibrates in the plane of the driving force. When the drive amplitude is increased, the wire leaves that plane and whirls, each point tracing
an ellipse around the line joining its supports. At still larger driving
amplitudes the whirl becomes modulated, and in a narrow range of
frequencies the direction of the whirl reverses at irregular intervals. All
of this can be seen with a strobe light, and it has been measured with
simple optical sensors~\cite{Hanson1994,Molteno1990,Molteno2004}.

The full equations for a string whose tension obeys Hooke's law are
nonlinear because of the geometry: the stretch of the string depends
nonlinearly on its displacement. These equations date back at least to
Kirchhoff~\cite{Kirchhoff1876}. Carrier reduced them to a wave equation
with a tension set by the total stretch of the string~\cite{Carrier1945}.
Murthy and Ramakrishna measured the nonlinear resonance of a driven wire,
including its non-planar motion, and Miles explained why the planar motion
loses stability~\cite{Murthy1965,Miles1965}. Narasimha derived the
equations of a flexible string with both transverse polarizations and
longitudinal motion, reduced them to the Kirchhoff--Carrier form when the
longitudinal waves are fast, and showed how the planar motion becomes
unstable to whirling~\cite{Narasimha1968}; Morse and Ingard and Anand gave
the corresponding cubic equations~\cite{Morse1968,Anand1969}, Tagata treated
the finite-amplitude parametric oscillation of a string driven along its
length~\cite{Tagata1977}, and
Antman set out the exact equations and their
approximations~\cite{Antman1980}. Miles then wrote the averaged equations
for the slow amplitudes of the two polarizations of the fundamental
mode~\cite{Miles1984}, and the averaged equations, with the two-mode equations from which
they come, became the standard models of the whirling string.

Chaos entered the story at the end of the 1980s. Johnson and Bajaj found
amplitude-modulated whirling and chaos in the averaged equations, and torus
doubling and chaos in the two-mode equations~\cite{Johnson1989}. Chaotic string vibrations, including torus
doubling, were also found in simulations and in experimental studies of a
vibrating wire~\cite{Tufillaro1989,Molteno1990}.
O'Reilly and Holmes compared experiments on a wire driven at one end with
the averaged equations. The predicted frequencies agreed well with the
measurements, but the predicted drive amplitudes were off by an order of
magnitude. O'Reilly and Holmes also studied the chaos in the limit of zero
damping, where the averaged equations are Hamiltonian~\cite{OReilly1992},
drawing on Mielke, Holmes and O'Reilly's analysis of chaos near homoclinic
orbits in such systems~\cite{Mielke1992}. Bajaj and Johnson followed the averaged equations to the crisis in which the chaotic attractor is destroyed~\cite{Bajaj1992}, and O'Reilly located in the damped, driven averaged equations the Shil'nikov homoclinic orbit near which the chaos at the lower damping arises~\cite{OReilly1993}.

Between 1992 and 2000 the string was
studied beyond the cubic approximation. Watzky~\cite{Watzky1992} and Gottlieb~\cite{Gottlieb1996} treated its large-amplitude, non-planar free vibration,
Rubin and Gottlieb integrated the whirling string numerically with the
longitudinal motion kept and the drive applied at the support, and still
found whirling and aperiodic motion at a drive about five times smaller
than in O'Reilly and Holmes' experiment~\cite{Rubin1996}; Bayly and Murphy showed that a
static load added to the drive breaks the symmetry of the string and couples
its second in-plane mode to the first out-of-plane mode~\cite{Bayly1998};
and Leamy and Gottlieb showed how the longitudinal dynamics and material
nonlinearity enter the whirl through internal resonances~\cite{Leamy2000}.

More recently, advances have been made in techniques for simulating string
motion. Energy-conserving schemes for nonlinear strings came from musical acoustics~\cite{Bilbao2005,Bilbao2006}; they are implicit, and for the exact nonlinearity they must be solved by a nonlinear iteration at each time step~\cite{Ducceschi2022}. Ducceschi and Bilbao replaced that iteration with energy
quadratization, a non-iterative method that needs only one linear solve per
time step~\cite{Ducceschi2022}, and the geometrically exact equations are
now a standard tool in sound synthesis~\cite{Bilbao2023,Russo2025}. 

The string is perhaps the most basic example of the dynamics of an elastic
body, yet the history of its models is convoluted; even the name and the
proper form of the string equation are still
discussed~\cite{Suchde2016,Suchde2017}. The models in common use are at
three levels of approximation: the partial differential equations of the
full string, the Kirchhoff--Carrier equation truncated to two modes, and a
four-dimensional averaged flow. Each is an approximation of the one
before it, so the three can be compared systematically. Such a comparison serves
two purposes. It shows where each approximation can be trusted. It also shows how an infinite-dimensional model and its
low-dimensional, typically modal, approximations are related when their
dynamics are chaotic.

The closest earlier work on the string compared the averaged equations
with the two-mode equations~\cite{Johnson1989,Bajaj1992} or with
experiment~\cite{Molteno2004}. Elsewhere in structural
dynamics, Holmes derived the one-mode Duffing model of a buckled beam from
the beam equation and compared it with Moon's
experiments~\cite{Holmes1979,Moon1979}, Cusumano and Moon did the same for
the non-planar chaos of a thin elastica~\cite{Cusumano1995a,Cusumano1995b},
and Galerkin and direct reductions of cables and beams have been compared~\cite{Pakdemirli1995,Nayfeh1997,Rega2004,Touze2006}. Comparisons
that go through a route to chaos are
rarer~\cite{Curry1984,Armbruster1989,Touze2011}, and none that we know of
follows how the parameter value at which each bifurcation occurs shifts
with the accuracy of the reduction.

In this paper we follow, for the whirling string, how each bifurcation on the route to chaos shifts from one model to the next. We integrate
three models under similar parameter conditions: the geometrically exact
(GE) string with two transverse polarizations and longitudinal motion, the
two-mode Kirchhoff--Carrier equations, and Miles' averaged equations. The
first two give the displacement of the string at each instant, including
its oscillation at the drive frequency; the averaged equations give only
the slowly varying amplitude and phase of that oscillation. We
locate the same bifurcation points in all
three, from the Hopf bifurcation of the steady whirl to the primary
homoclinic orbit, and measure how they shift with $\eps = c/\omega_1$, the
ratio of the slow rate $c$ at which the amplitude and phase of the whirl
change to the linear fundamental frequency $\omega_1$. Averaging is
expected to be accurate when this ratio is small (Sec.~\ref{sec:leaves}). In our computations $\eps$ ranges from 0.005 to 0.02. We first compare the
averaged equations with the two-mode equations they come from---which
isolates the error of averaging---and then compare the two-mode equations
with the full string---which measures both the modal truncation and the
Kirchhoff--Carrier approximation. To separate an error that grows with $\eps$ from one that grows with
amplitude, we repeat some runs with a stiffer string, which whirls at a
smaller amplitude for the same $\eps$.

We find that all three models show the same route to chaos, and that the
averaged equations shift local bifurcations by an amount of order $\eps$, consistent with the order-$\eps$ shift that averaging theory gives for the Hopf bifurcation and leads one to expect for the period doublings (Sec.~\ref{sec:leaves}). The GE and two-mode models agree with each other to 0.002 in the scaled detuning at the local bifurcations for $\eps \le 0.01$ (one period halving, uncertain by 0.003, differs by 0.004), so most of the difference between the averaged
equations and the full string comes from the averaging. For the global bifurcations no theorem gives the size of the error. Global bifurcations can be shifted by several times $\eps$, and the size of this error is set by how
steeply the splitting function of the homoclinic loop, defined in
Sec.~\ref{sec:terms}, crosses zero. The
study is offered, in part, as an example of what to expect from a reduced model
near a global bifurcation.

Section~\ref{sec:terms} reviews the concepts from nonlinear dynamics that
we need, Sec.~\ref{sec:models} sets out the models, and
Sec.~\ref{sec:methods} the parameters and numerical methods.
Section~\ref{sec:qual} shows where the models agree,
Sec.~\ref{sec:steps} where they differ, and Sec.~\ref{sec:hom} explains why
the homoclinic bifurcation is the most sensitive to the modeling approximations.

% ============================================================================
\section{Review of concepts from nonlinear dynamics}
\label{sec:terms}

This study relies on concepts from nonlinear dynamics that may be unfamiliar to some readers. To make the paper
self-contained, we begin with a brief review. Standard introductions to the details can be found in the texts of Thompson
and Stewart~\cite{Thompson2002}, Nayfeh and
Balachandran~\cite{NayfehBalachandran1995} and Moon~\cite{Moon1992}, which
are written for the vibration reader, and in Strogatz's more elementary
book~\cite{Strogatz2015}; Guckenheimer and Holmes~\cite{Guckenheimer1983},
Wiggins~\cite{Wiggins2003}, Glendinning~\cite{Glendinning1994} and
Kuznetsov~\cite{Kuznetsov2004} give the theory.

The averaged equations of Sec.~\ref{sec:models} are a four-dimensional
autonomous flow for the slow amplitudes of the two polarizations. A steady
whirl of the string is an equilibrium of this flow. The stability of the steady whirl is determined by the eigenvalues of the Jacobian, and when a complex pair crosses the
imaginary axis the equilibrium gives birth to a small periodic orbit: a
\emph{Hopf bifurcation}. For the string this is the point at which the
amplitude of the steady whirl begins to oscillate slowly. In the two-mode and GE models the steady
whirl is not an equilibrium but a periodic orbit at the drive frequency,
and the same event is a \emph{Neimark--Sacker} bifurcation of that orbit,
located by the \emph{Floquet multipliers} of the map that advances the
state by one drive period (the stroboscopic map); a complex pair of
multipliers leaves the unit circle. Since the Hopf and Neimark--Sacker bifurcations are the same physical
event, we call both the Hopf bifurcation of the whirl. A \emph{period doubling} is the point at which a
periodic orbit of the slow flow loses stability and gives birth to an orbit
of twice its period;
in the string the modulation of the whirl alternates between two
amplitudes. In the averaged equations the period doubling is located by a Floquet multiplier of the
Poincar\'e return map passing through $-1$. In the GE and two-mode models the
modulated whirl is a torus in the phase space of the forced system (the
drive period and the modulation period are the two angles), and the same
event is a \emph{torus doubling}. A \emph{period-doubling
cascade} is a common route to chaos, and the chaotic attractor it produces is
interrupted by \emph{periodic windows}, intervals of parameter in which a
periodic orbit is again attracting.

The chaos of the whirling string is organized by a mechanism known as
\emph{Shil'nikov chaos}. On
the upper branch of the string's resonance curve the planar motion is an
equilibrium of the slow flow with one unstable direction (out of the
plane), a stable spiral within the plane, and a fourth direction, also out
of the plane, that is stable and decays faster than the spiral: a
\emph{saddle focus}, which is the case Shil'nikov's theorem covers. An orbit that
leaves this equilibrium along its unstable direction and returns to it along
the stable spiral is a \emph{homoclinic orbit}, and the parameter value at
which it exists is a \emph{homoclinic bifurcation}. Unlike the Hopf bifurcation and the
period doublings, which are \emph{local} bifurcations found from the
eigenvalues or Floquet multipliers of a single equilibrium or periodic orbit,
a homoclinic bifurcation is \emph{global}: it depends on the whole of an
orbit that travels far from the equilibrium.

Shil'nikov proved that if the spiral's decay rate
$|\rho|$ is smaller than the unstable rate $\lambda_u$, then arbitrarily
near the homoclinic orbit there are infinitely many periodic orbits and
horseshoes, which is to say chaotic invariant
sets~\cite{Shilnikov1965,Glendinning1984,Gaspard1984}; if $|\rho| > \lambda_u$
there are not. We call $|\rho|/\lambda_u$ the \emph{Shil'nikov ratio}. In the averaged equations the in-plane decay rate is the scaled damping $\gamma$ of Eq.~(\ref{eq:scalings}), so $\rho = -\gamma$ and the Shil'nikov ratio is $\gamma/\lambda_u$. The same holds in the two-mode equations: the damping is the same on both in-plane variables, so the in-plane Floquet exponents of the planar state sum to $-2\gamma$ in slow time, and when they form a complex pair each has real part $-\gamma$. For the GE string we measure $\rho$ (Sec.~\ref{sec:protocol}). An orbit near the saddle focus
spirals slowly in, is then thrown out along the unstable direction, makes a
large excursion (a burst of whirling motion), and comes back to
spiral in again; the number of turns it makes before leaving depends
sensitively on how close it came, and this sensitivity is the source of the chaos. To
locate the homoclinic bifurcation we use the \emph{splitting function}
$\beta(\Delta)$, where $\Delta$ is the scaled detuning of Sec.~\ref{sec:models}: the
signed distance, at the first return, between the unstable manifold leaving
the saddle and the stable manifold entering it. The zero of $\beta$ is the homoclinic
orbit. We take $\beta$ from Kuznetsov, who calls it the split function and takes its zero as the condition for a homoclinic bifurcation~\cite[Sec.~6.1]{Kuznetsov2004}; in
most of his homoclinic bifurcation theorems a nonzero slope at that zero is
one of the genericity conditions~\cite[Secs.~6.2--6.3]{Kuznetsov2004}. To
first order the slope is given by a Melnikov integral along the homoclinic
orbit~\cite[Sec.~6.4.1]{Kuznetsov2004}, the general form of the Melnikov
function familiar from the analysis of forced
oscillators~\cite{Guckenheimer1983,Wiggins2003}. This slope decides how sensitive the homoclinic bifurcation is to the modeling approximations (Sec.~\ref{sec:hom}).

Three more terms appear. A \emph{crisis} is the sudden destruction of a
chaotic attractor when it collides with an unstable orbit on the boundary of
its basin~\cite{Grebogi1983}; here the crisis is where the chaotic whirling ends
and the string falls back to planar motion. The largest \emph{Lyapunov exponent} is the mean
rate at which two nearby trajectories separate; it is positive on a
chaotic attractor, and its reciprocal is the time over which the memory of
the initial state is lost. A \emph{first-return map}
records one coordinate at successive crossings of a surface in phase space
and plots each value against the previous one; for a strongly dissipative
system the points fall nearly on a curve, and the shape of that curve is a
fingerprint of the attractor. Finally, \emph{averaging} is the method,
familiar to vibration engineers from Nayfeh and Mook~\cite{NayfehMook1979},
of replacing a weakly nonlinear oscillation at the drive frequency by
equations for its slowly varying amplitude and phase; the result is correct
to first order in the ratio of the slow rate to the drive frequency, which
is the small parameter $\eps$ of this paper~\cite{Sanders2007}.

% ============================================================================
\section{Three models of a whirling string}
\label{sec:models}

\subsection{The geometrically exact string}

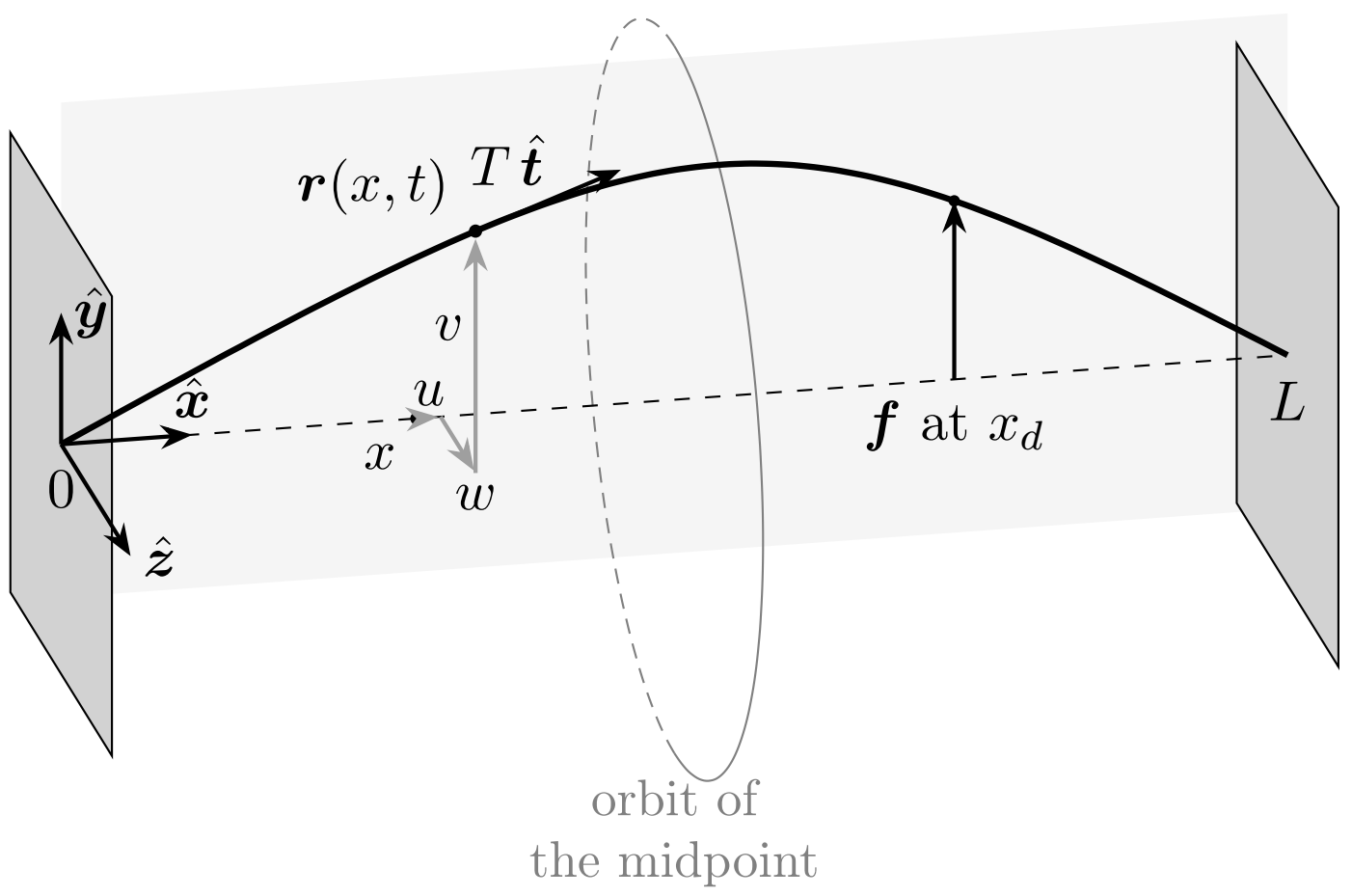
\begin{figure}[t]
\centering
\begin{tikzpicture}[scale=0.86,>={Stealth[length=2.2mm]},font=\small]
\fill[gray!8] (0.000,-1.222) -- (9.781,-0.511) -- (9.781,3.436) -- (0.000,2.725) -- cycle;
\fill[gray!35, draw=black, thin] (-0.405,-1.180) -- (-0.405,2.485) -- (0.405,1.180) -- (0.405,-2.485) -- cycle;
\fill[gray!35, draw=black, thin] (9.376,-0.469) -- (9.376,3.196) -- (10.187,1.891) -- (10.187,-1.774) -- cycle;
\draw[gray, thin, densely dashed] plot[smooth] coordinates {(4.891,-2.464) (4.854,-2.400) (4.817,-2.329) (4.780,-2.251) (4.744,-2.165) (4.708,-2.073) (4.672,-1.974) (4.637,-1.869) (4.603,-1.757) (4.570,-1.640) (4.537,-1.517) (4.506,-1.389) (4.475,-1.257) (4.446,-1.119) (4.418,-0.978) (4.391,-0.834) (4.365,-0.685) (4.341,-0.535) (4.319,-0.381) (4.298,-0.226) (4.279,-0.069) (4.261,0.089) (4.245,0.248) (4.231,0.407) (4.218,0.566) (4.208,0.725) (4.199,0.882) (4.193,1.038) (4.188,1.192) (4.185,1.344) (4.184,1.493) (4.185,1.639) (4.188,1.781) (4.193,1.920) (4.199,2.054) (4.208,2.184) (4.218,2.308) (4.231,2.428) (4.245,2.541) (4.261,2.649) (4.279,2.750) (4.298,2.845) (4.319,2.933) (4.341,3.014) (4.365,3.087) (4.391,3.153) (4.418,3.212) (4.446,3.262) (4.475,3.305) (4.506,3.339) (4.537,3.366) (4.570,3.384) (4.603,3.394) (4.637,3.395) (4.672,3.388) (4.708,3.373) (4.744,3.350) (4.780,3.318) (4.817,3.278) (4.854,3.230) (4.891,3.175)};
\draw[gray, thin] plot[smooth] coordinates {(4.891,3.175) (4.928,3.111) (4.965,3.040) (5.001,2.962) (5.038,2.877) (5.074,2.784) (5.109,2.685) (5.144,2.580) (5.178,2.468) (5.212,2.351) (5.244,2.228) (5.276,2.100) (5.306,1.968) (5.336,1.831) (5.364,1.689) (5.391,1.545) (5.416,1.397) (5.440,1.246) (5.463,1.092) (5.484,0.937) (5.503,0.780) (5.521,0.622) (5.537,0.463) (5.551,0.304) (5.563,0.145) (5.574,-0.014) (5.582,-0.171) (5.589,-0.327) (5.594,-0.481) (5.597,-0.633) (5.598,-0.782) (5.597,-0.928) (5.594,-1.070) (5.589,-1.209) (5.582,-1.343) (5.574,-1.473) (5.563,-1.597) (5.551,-1.717) (5.537,-1.830) (5.521,-1.938) (5.503,-2.039) (5.484,-2.134) (5.463,-2.222) (5.440,-2.303) (5.416,-2.376) (5.391,-2.442) (5.364,-2.501) (5.336,-2.551) (5.306,-2.594) (5.276,-2.628) (5.244,-2.655) (5.212,-2.673) (5.178,-2.682) (5.144,-2.684) (5.109,-2.677) (5.074,-2.662) (5.038,-2.638) (5.001,-2.607) (4.965,-2.567) (4.928,-2.519) (4.891,-2.464)};
\draw[dashed] (0.000,0.000) -- (9.781,0.711);
\draw[very thick] plot[smooth] coordinates {(0.000,0.000) (0.121,0.067) (0.242,0.134) (0.363,0.200) (0.484,0.267) (0.605,0.333) (0.725,0.399) (0.845,0.464) (0.965,0.529) (1.084,0.594) (1.202,0.658) (1.320,0.722) (1.438,0.784) (1.555,0.847) (1.671,0.908) (1.786,0.968) (1.900,1.028) (2.014,1.087) (2.127,1.144) (2.239,1.201) (2.350,1.256) (2.460,1.311) (2.569,1.364) (2.678,1.416) (2.785,1.467) (2.891,1.516) (2.996,1.564) (3.100,1.611) (3.203,1.656) (3.305,1.700) (3.406,1.742) (3.506,1.783) (3.606,1.822) (3.704,1.859) (3.801,1.895) (3.897,1.929) (3.992,1.961) (4.086,1.992) (4.180,2.021) (4.272,2.048) (4.364,2.074) (4.455,2.097) (4.545,2.119) (4.634,2.139) (4.723,2.158) (4.811,2.174) (4.899,2.189) (4.986,2.201) (5.072,2.212) (5.158,2.221) (5.244,2.228) (5.330,2.234) (5.415,2.237) (5.499,2.239) (5.584,2.238) (5.668,2.236) (5.753,2.232) (5.837,2.227) (5.921,2.219) (6.006,2.210) (6.090,2.199) (6.175,2.187) (6.259,2.172) (6.344,2.156) (6.429,2.139) (6.515,2.119) (6.600,2.098) (6.686,2.076) (6.773,2.052) (6.860,2.026) (6.947,1.999) (7.035,1.971) (7.123,1.941) (7.212,1.910) (7.301,1.877) (7.390,1.843) (7.481,1.808) (7.571,1.772) (7.663,1.734) (7.755,1.696) (7.847,1.656) (7.940,1.615) (8.033,1.574) (8.127,1.531) (8.222,1.487) (8.316,1.443) (8.412,1.398) (8.508,1.352) (8.604,1.305) (8.701,1.258) (8.798,1.210) (8.895,1.162) (8.993,1.113) (9.091,1.064) (9.189,1.014) (9.287,0.964) (9.386,0.914) (9.485,0.863) (9.584,0.813) (9.682,0.762) (9.781,0.711)};
\fill (2.837,0.206) circle (1.5pt) node[below left=1pt and 0pt] {$x$};
\draw[->, thick, gray!75, text=black, shorten >=0.5pt] (2.837,0.206) -- (3.026,0.220) node[midway, above, inner sep=2pt] {$u$};
\draw[->, thick, gray!75, text=black, shorten >=0.5pt] (3.026,0.220) -- (3.305,-0.229) node[pos=1, below, inner sep=2pt] {$w$};
\draw[->, thick, gray!75, text=black, shorten >=1.5pt] (3.305,-0.229) -- (3.305,1.700) node[pos=0.6, left, inner sep=2pt] {$v$};
\fill (3.305,1.700) circle (1.5pt);
\node[above left, inner sep=3pt, xshift=-2pt] at (3.305,1.700) {$\bm r(x,t)$};
\draw[->, thick] (3.305,1.700) -- (4.465,2.190) node[pos=0.55, above left, inner sep=2pt] {$T\,\hat{\bm t}$};
\draw[->, thick] (7.123,0.518) -- (7.123,1.941);
\node[below, inner sep=4pt] at (7.123,0.518) {$\bm f$ at $x_d$};
\fill (7.123,1.941) circle (1.3pt);
\node[gray, font=\footnotesize, align=center, below, inner sep=5pt] at (4.891,-2.464) {orbit of\\the midpoint};
\draw[->, thick] (0,0) -- (1.047,0.076) node[above, inner sep=3pt] {$\hat{\bm x}$};
\draw[->, thick] (0,0) -- (0.000,1.050) node[right, inner sep=2pt] {$\hat{\bm y}$};
\draw[->, thick] (0,0) -- (0.554,-0.892) node[right, inner sep=2pt] {$\hat{\bm z}$};
\node[below, inner sep=5pt] at (0.000,0.000) {$0$};
\node[below, inner sep=5pt] at (9.781,0.711) {$L$};
\end{tikzpicture}
\caption{Geometry of the string. The material point $x$ on the stretched
equilibrium (dashed) is displaced to $\bm r = (x+u)\,\hat{\bm x} +
v\,\hat{\bm y} + w\,\hat{\bm z}$, and the tension $T$ acts along the unit
tangent $\hat{\bm t}$. The drive $\bm f$ acts in the $y$ direction at
$x = x_d$, in the shaded $x$--$y$ plane. In a steady whirl each point
traces an ellipse about the $x$ axis; the ellipse of the midpoint is drawn in
grey. The arrows at the marked point show the displacement of $x$ resolved
into its longitudinal part $u$ and its transverse parts $v$ and $w$. The displacements are exaggerated.}
\label{fig:geom}
\end{figure}

Label each material point of the string by its position $x$ along the
stretched equilibrium, $0 \le x \le L$, with tension $T_0$ and mass per unit
length $\mu$. At time $t$ the point $x$ is at
$\rr = (x + u)\,\hat{\bm x} + v\,\hat{\bm y} + w\,\hat{\bm z}$, where $u$ is
the longitudinal displacement and $v$ and $w$ are the transverse ones, $v$
in the plane of the drive and $w$ out of it (Fig.~\ref{fig:geom}).
The local stretch is $\nu = |\partial_x\rr|$, the unit tangent is
$\that = \partial_x\rr/\nu$, and for a perfectly flexible string the tension is $T = T_0 + K(\nu - 1)$ and acts along $\that$, where $K$, Young's modulus times the cross-sectional area, is the axial stiffness. Newton's second law for an element then gives
\begin{equation}
\mu\,\partial_t^2\rr = \partial_x\bigl(T\,\that\bigr) - 2\mu\sigma\,
\partial_t\rr_\perp + \bm f(x,t) ,
\label{eq:GE}
\end{equation}
where $\rr_\perp = v\,\hat{\bm y} + w\,\hat{\bm z}$, $\sigma$ is the damping rate, so that the amplitude of each free transverse mode decays as $e^{-\sigma t}$, and $\bm f$ is the external force per unit length. Equation~(\ref{eq:GE})
holds at any amplitude. It treats the two transverse directions alike, so
that the angular momentum of the string about the $x$ axis changes only
through the external force and the damping. In the computations a small
bending stiffness is added to the transverse equations, and the
longitudinal motion carries its own light damping. The energy can be written
as the linear wave energy at tension $T_0$ plus the non-negative term
$\frac12(K - T_0)\int(\nu - 1)^2\,dx$, which holds all of the
nonlinearity~\cite{Ducceschi2022}; this form is what makes the numerical
scheme of Sec.~\ref{sec:methods} possible.

\subsection{The two-mode Kirchhoff--Carrier equations}

When the tension is small compared with the axial stiffness, $T_0 \ll K$, the longitudinal waves are much faster than the transverse ones; for the steel wire used below their speed ratio
$\sqrt{K/T_0}$ is about 36. The longitudinal motion then adjusts at once to
the transverse shape, the tension becomes uniform along the string, and its
value is set by the total stretch. This is the Kirchhoff--Carrier
approximation~\cite{Kirchhoff1876,Carrier1945}. Projecting onto the first spatial Fourier mode in each transverse plane (a one-mode Galerkin projection), $v = v_1(t)\sin(\pi x/L)$ and $w = w_1(t)\sin(\pi x/L)$, and driving with a point force $F\cos\Omega t$ in the $y$ direction at
$x = x_d$, gives
\begin{equation}
\begin{aligned}
\ddot v_1 + 2\sigma\dot v_1 + \omega_1^2\bigl[1 + g\,(v_1^2 + w_1^2)\bigr]\,v_1
 &= F_d\cos\Omega t ,\\
\ddot w_1 + 2\sigma\dot w_1 + \omega_1^2\bigl[1 + g\,(v_1^2 + w_1^2)\bigr]\,w_1
 &= 0 ,
\end{aligned}
\label{eq:twomode}
\end{equation}
where $v_1$ and $w_1$ are the displacements of the midpoint, $\omega_1$ is the linear
fundamental, $F_d = (2F/\mu L)\sin(\pi x_d/L)$, and
$g = (K/4T_0)(\pi/L)^2$. Each line is a hardening Duffing oscillator, and the
two are coupled through the squared distance $v_1^2 + w_1^2$ of the midpoint
from the axis. Without damping and drive, Eqs.~(\ref{eq:twomode}) describe a
particle in a central potential, and the circular whirl is an exact
solution~\cite{Elliott1980,Tufillaro1989}.

\subsection{Miles' averaged equations}

When the damping, the detuning, the nonlinear frequency shift and the drive each change the motion by only a small fraction during one period, the motion is a slowly modulated oscillation at the drive frequency. Write $v_1 = s\,\mathrm{Re}[V\exp(i\Omega t)]$ and
$w_1 = s\,\mathrm{Re}[W\exp(i\Omega t)]$, with slowly varying complex
amplitudes $V$ and $W$, introduce the slow time
$\tau = ct$, and average Eqs.~(\ref{eq:twomode}) over one drive period (the rotating-wave approximation of optics). With
the scalings
\begin{equation}
\gamma = \frac{\sigma}{c},\qquad
\Delta = \frac{\Omega^2 - \omega_1^2}{2\Omega c},\qquad
\alpha = \frac{F_d}{2\Omega s c},\qquad
s^2 = \frac{4\Omega c}{\omega_1^2 g},
\label{eq:scalings}
\end{equation}
the averaged equations are
\begin{equation}
\begin{aligned}
V' &= -\gamma V - i\bigl(\Delta - \tfrac32 E\bigr) V + J W - i\alpha ,\\
W' &= -\gamma W - i\bigl(\Delta - \tfrac32 E\bigr) W - J V ,
\end{aligned}
\label{eq:avg}
\end{equation}
where the prime is $d/d\tau$, $E = |V|^2 + |W|^2$ is the scaled
energy, and $J = \mathrm{Im}(W^* V)$ is the scaled angular momentum
about the axis. The term $\frac32 E$ is the amplitude-dependent frequency
shift of the Duffing oscillator, and the terms $JW$ and $-JV$ rotate energy
between the two polarizations. With $V = p_1 - iq_1$ and $W = p_2 - iq_2$ Eqs.~(\ref{eq:avg}) are Miles' equations~\cite{Miles1984}; in O'Reilly's notation they have the coefficients $A = -3/2$ and $B = 1$ that belong to the Kirchhoff--Carrier string~\cite{OReilly1993}.
Only the combinations $\gamma/\alpha^{2/3}$ and $\Delta/\alpha^{2/3}$
matter, so we set $\alpha = 1$; the choice of $c$ is then free, and it fixes
how the averaged parameters $(\gamma, \Delta)$ map onto a physical damping
and drive frequency.

Equations~(\ref{eq:avg}) are invariant under $W \to -W$, the reflection
$z \to -z$ of the string, so whirling in either direction is equally
possible. Planar motions, $W = 0$, have energies given by the cubic
\begin{equation}
\tfrac94 E^3 - 3\Delta E^2 + (\gamma^2 + \Delta^2)E - 1 = 0 ,
\label{eq:planar}
\end{equation}
which is the familiar frequency-response curve of the hardening Duffing oscillator, bent toward higher frequency, with its jumps and hysteresis~\cite{NayfehMook1979}; see also Ref.~\cite[Secs.~3.5 and 3.7]{Tufillaro1992} for the string, including the regions of planar and whirling motion. Linearizing about a planar state
gives two in-plane eigenvalues
$-\gamma \pm \sqrt{-(\Delta - \frac32E)(\Delta - \frac92E)}$ and two
out-of-plane eigenvalues $-\gamma \pm \sqrt{-(\Delta - \frac32E)(\Delta - \frac12E)}$.
On the upper branch of the resonance curve the in-plane pair is complex,
$-\gamma \pm i\omega$, and over a range of $\Delta$ one out-of-plane
eigenvalue $\lambda_u$ is positive: the planar state is a saddle focus. Because $\lambda_u$ is positive, a small out-of-plane disturbance grows: the orbit leaves the planar state along the unstable direction, and the string begins to whirl. If the orbit returns to the planar state along the stable spiral, it is a Shil'nikov homoclinic orbit~\cite{Shilnikov1965}. The complex pair decays more slowly than the third stable eigenvalue, so an orbit returning to the planar state spirals in, and Shil'nikov's theorem applies: if $\gamma/\lambda_u < 1$ there are infinitely many
periodic orbits near the homoclinic orbit and horseshoes; if $\gamma/\lambda_u > 1$ there are
not~\cite{Glendinning1984,Gaspard1984,Guckenheimer1983,Kuznetsov2004}. Whether
the chaotic sets are attracting is a separate question, answered here by
computation.
O'Reilly located the Shil'nikov homoclinic orbit in Eqs.~(\ref{eq:avg}) at
$\Delta^* = 3.078$ for $\gamma = 0.45$ and at $\Delta^* = 3.405$ for
$\gamma = 0.50$, on either side of $\gamma/\lambda_u = 1$~\cite{OReilly1993}.
We work at these two values of $\gamma$.

\subsection{Approximations of the geometrically exact string}
\label{sec:leaves}

\begin{figure}[t]
\centering
\begin{tikzpicture}[
    box/.style={draw, rounded corners, align=flush center, text width=0.95\textwidth,
              inner sep=6pt, font=\footnotesize},
  arr/.style={-{Stealth[length=3mm]}, thick},
  lab/.style={font=\footnotesize, align=left, right=6pt, text width=0.95\textwidth-1.6cm}]
\node[box] (ge) {\textbf{Geometrically exact string} (PDE, Eq.~\ref{eq:GE}): $u$, $v$, $w$\\
  tension along the tangent; exact stretch\\
  Kirchhoff (1883), Narasimha (1968), Antman (1980); numerics:\\
  Bilbao (2005, 2006), Rubin and Gottlieb (1996), Ducceschi and Bilbao (2022)};
\node[box, below=1.6cm of ge] (kc) {\textbf{Two-mode Kirchhoff--Carrier equations}
  (ODE, Eq.~\ref{eq:twomode}): $v_1(t)$, $w_1(t)$\\
  Carrier (1945), Miles (1965), Narasimha (1968), Elliott (1980), Tufillaro (1989)\\
  Johnson and Bajaj (1989), O'Reilly and Holmes (1992)};
\node[box, below=1.6cm of kc] (av) {\textbf{Averaged equations}
  (slow flow, Eq.~\ref{eq:avg}): $V(\tau)$, $W(\tau)$\\
  Miles (1984), Johnson and Bajaj (1989),\\
  Bajaj and Johnson (1992), O'Reilly (1993)};
\draw[arr] ([xshift=1.2cm]ge.south west) -- ([xshift=1.2cm]kc.north west)
  node[midway, lab] {Steps A and B: fast longitudinal waves ($T_0/K \ll 1$);\\
  quadratic stretch; one mode per plane; no bending stiffness};
\draw[arr] ([xshift=1.2cm]kc.south west) -- ([xshift=1.2cm]av.north west)
  node[midway, lab] {Step C: first-order averaging over one drive period;\\
  error $O(\eps)$, with $\eps = c/\omega_1$};
\end{tikzpicture}
\caption{The three models compared in this paper and the approximations
that lead from one to the next, with representative references. The middle model is the Galerkin truncation of the Kirchhoff--Carrier string to its first spatial mode in each of the two transverse planes, the two-mode model; the bottom one is its first-order average.}
\label{fig:ladder}
\end{figure}
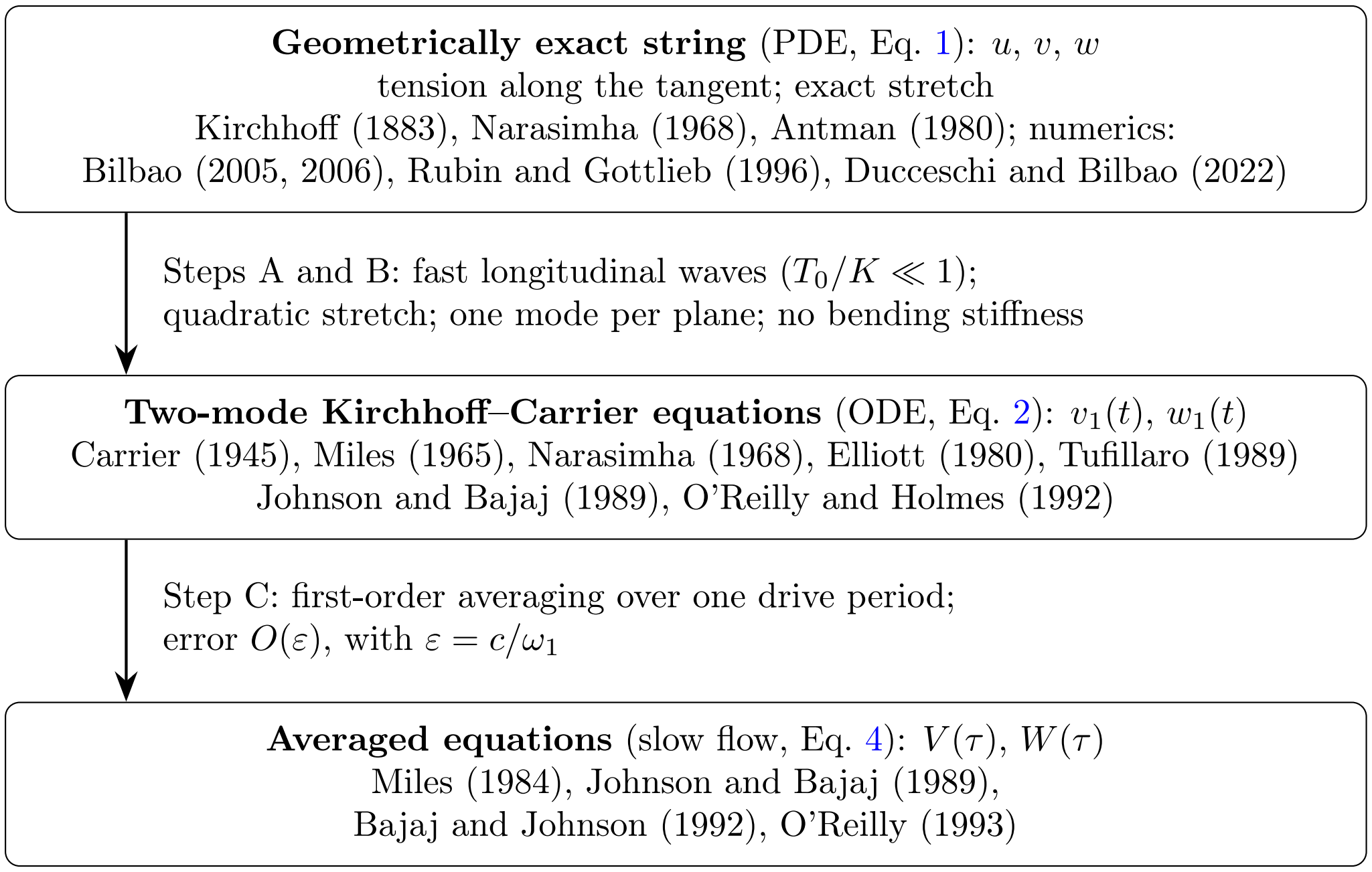

Figure~\ref{fig:ladder} summarizes the three models and the approximations
that connect them. We group the approximations into three steps, A, B and C.

Step A is the Kirchhoff--Carrier approximation. The GE equations
(\ref{eq:GE}) carry the longitudinal displacement $u$ as a dynamical
variable; the Kirchhoff--Carrier equation replaces it by its quasi-static
value, so that the strain is uniform along the string and the tension is
set by the total stretch. Step A neglects the dynamics of the longitudinal
waves, whose lowest frequency is about 2.5~kHz for the wire used below, and
terms of relative size $T_0/K$, below $10^{-3}$ here. Step A also replaces the
exact stretch $\nu - 1$ by its quadratic approximation in the slopes.
None of the approximations in step A has a small parameter that we vary; all are controlled by
the amplitude. In the computations the GE string also carries a small
bending stiffness, so that its linear fundamental differs slightly from the
flexible one; the drive is referenced to each model's own fundamental, so
this difference does not enter the comparison.

Step B is the modal truncation. Keeping one mode in each polarization neglects the higher transverse modes, which are not resonant with the drive and
enter only as small corrections to the effective cubic coefficient. Steps A
and B cannot be separated in our runs, because the two-mode equations
(\ref{eq:twomode}) are the Kirchhoff--Carrier string truncated to one mode;
we measure their combined error by comparing the two-mode string with the GE
string.

Step C is the averaging. It neglects terms of relative size
\begin{equation}
\eps = \frac{c}{\omega_1} ,
\label{eq:eps}
\end{equation}
which is the ratio of the slow rate to the fast one. By the averaging theorem, a solution of the averaged equations stays within a distance of order $\eps$ of the corresponding solution of the two-mode equations for slow times $\tau$ of order one, that is, for times $t$ of order $1/c$~\cite{Sanders2007}. In physical terms $\eps = \sigma/(\gamma\,\omega_1)$, the damping rate divided by $\gamma\,\omega_1$. Since $\Delta \approx (\Omega - \omega_1)/c$,
an error $\delta\Delta$ in the scaled detuning is an error
$c\,\delta\Delta$ in the drive frequency. We measure the error of step C by
comparing the averaged equations with the two-mode string as we vary $\eps$ by a factor of four, through $\sigma = 1$, 2 and 4~s$^{-1}$.

We ask how far the terms neglected in each step move the bifurcation points along the route to chaos. Near a hyperbolic equilibrium or periodic orbit of the averaged flow, averaging theory bounds the error for all times~\cite{Sanders2007}, and it places the Hopf bifurcation of the string within order $\eps$ of that of the averaged flow; for the period doublings of the modulated whirl, which are torus doublings of the string, the same is expected but not proved. Along a route to chaos the attractors are not hyperbolic, and, as far as we know, no theorem gives the size of the error for them or for the global bifurcations, so the error has to be measured. In the simulations used below the whirl amplitude grows with the damping as well as with $\eps$ (Sec.~\ref{sec:params}), so a difference that grows with $\sigma$ could come from step A or B, through the amplitude, or from step C, through $\eps$. A separate set of simulations, with the axial stiffness $K$ scaled by 2 and 4, keeps $\eps$ at its value for $\sigma = 2$ and 4~s$^{-1}$ but returns the whirl amplitude to its value for $\sigma = 1$~s$^{-1}$, and so separates the effect of the amplitude (steps A and B) from the effect of $\eps$ (step C). These simulations show that the departure of the GE string from the two-mode string at the largest damping is mostly an amplitude effect (Sec.~\ref{sec:stepAB}).

% ============================================================================
\section{Parameters, scalings, and numerical methods}
\label{sec:methods}
\label{sec:params}

\subsection{The string}

All computations use a steel wire of length $L = 1$~m, radius 0.29~mm,
density 8000~kg/m$^3$, Young's modulus $2\times10^{11}$~Pa, and tension
$T_0 = 40$~N, driven in the $y$ direction at $x_d = 0.72L$. Its linear
fundamental is $f_1 = 68.78$~Hz and its cubic coefficient is
$g \approx 3.26\times10^3$~m$^{-2}$. For a chosen damping $\sigma$ and a point
$(\gamma, \Delta)$ of the averaged equations, Eqs.~(\ref{eq:scalings}) give
$c = \sigma/\gamma$, the drive frequency, the amplitude scale $s$, and the
drive force. For example, $\sigma = 1$~s$^{-1}$ and $\gamma = 0.45$ give
$c = 2.22$~s$^{-1}$, $s = 2.5$~mm, a drive force of about 7~mN, and
$\eps = 0.0051$; the chaotic band, from the first whirl-reversing
attractors near $\Delta = 3.0$ to the crisis near 3.21, then lies near
$\Omega/\omega_1 = 1.016$ and is about 0.07~Hz wide. We use $\sigma = 1$, 2 and
4~s$^{-1}$, which at $\gamma = 0.45$ give $\eps = 0.0051$, 0.0103 and
0.0206, and at $\gamma = 0.50$ give $\eps = 0.0046$, 0.0093 and 0.0185.
The same $\sigma$ is applied to every transverse mode, so the damping does
not distinguish the models.

\subsection{Methods for integrating the models}

The averaged equations are integrated with a fourth-order Runge--Kutta
method with step $0.01$ in $\tau$. The two-mode equations are integrated
with the same method at 60 steps per linear period. At this step the method lowers the linear frequency of the two-mode string by about $9\times10^{-7}$ of itself at $\sigma = 1$~s$^{-1}$ (less at the larger dampings, where the step is a smaller fraction of the drive period), so that the nominal values of $\Delta$ are low by about $9\times10^{-7}/\eps$, or 0.00017 at $\sigma = 1$~s$^{-1}$. Two-mode values of $\Delta$ are corrected for this offset, as the GE values are corrected below for the offset of their scheme, and the Floquet multipliers of the two-mode string (Sec.~\ref{sec:protocol}) are computed at 512 steps per drive period, where the offset is negligible. The GE string is
integrated with the energy-quadratised scheme of Ducceschi and
Bilbao~\cite{Ducceschi2022}, which we have extended from one transverse polarization to two (code in Ref.~\cite{Tufillaro2026data}). The nonlinear part of the energy is written as the
square of an auxiliary variable that is advanced with the displacements, so
that each step requires one linear solve and a discrete form of the energy, defined on the grid and the time steps, is conserved exactly in the absence of damping and drive~\cite{Yang2016,Shen2018}. The
transverse motion is discretized by finite differences (73 intervals, the finest grid that is stable at a sample rate of 16~kHz) and the longitudinal motion by its modes up to
5~kHz, of which there are two. The scheme includes bending stiffness, as needed to model real wires such as piano strings~\cite{Ducceschi2022}; for our wire it raises the linear fundamental by 0.014\% (0.009~Hz). Because the drive frequency is set relative to each model's own linear fundamental, this shift does not enter the comparison. A calibration run shows that the
fundamental of the discrete GE scheme lies 0.00096~Hz above the analytic
stiff-string value, which shifts the GE detuning by $0.0060/c$; all GE values
of $\Delta$ quoted below are corrected for the scheme's frequency offset. The effective cubic
coefficient of the GE string, measured from free circular whirls of 1 to
4~mm, agrees with the Kirchhoff--Carrier value to 0.1\%.

To compare the GE and two-mode models with Eqs.~(\ref{eq:avg}) we demodulate them to
the same slow amplitudes. For the GE string we take $v_1$ and $w_1$ to be the amplitudes of the
$\sin(\pi x/L)$ components of $v$ and $w$. From each amplitude and its
velocity we form $V = i\,(v_1 - i\dot v_1/\Omega)\exp(-i\Omega t)/s$ and
likewise $W$ from $w_1$ (the
factor $i$ because the computations drive the string with $\sin\Omega t$
rather than the $\cos\Omega t$ of Eq.~(\ref{eq:twomode}), a shift of the
time origin), and average over one drive period to remove the ripple at
$2\Omega$. The two-mode amplitudes are treated in the same way. The
``Hopf bifurcation'' and ``period doubling'' of the whirl are, in the GE and two-mode
models, a Neimark--Sacker bifurcation and a torus doubling of the periodic
whirl orbit, which the demodulation turns into the Hopf bifurcation and
period doubling of the slow amplitudes; we use the slow-flow names
throughout.

\subsection{Methods for locating the bifurcation points}
\label{sec:protocol}

Each run starts at the steady-whirl equilibrium of Eqs.~(\ref{eq:avg}) at
the same $(\gamma, \Delta)$, mapped to the physical state of each model,
runs for 300 units of $\tau$ ($300/c$, about 135~s at $\sigma = 1$~s$^{-1}$ and $\gamma = 0.45$) to let transients decay, and records the next
300 units. We scan $\Delta$ in steps of 0.0025 in all three models, from $\Delta = 2.62$ to 3.25 at $\gamma = 0.45$ and from 2.60 to 3.30 at $\gamma = 0.50$. Each attractor
is classified as a steady whirl, a periodic slow flow of period $k$ (counted by returns to the plane
$\mathrm{Re}\,V = \langle\mathrm{Re}\,V\rangle$),
or non-periodic, and as whirl-reversing if the angular momentum $J$ changes sign.

The local bifurcations are located in three ways. For the averaged and two-mode equations we locate them from the eigenvalues or Floquet multipliers of the linearization. For the averaged equations the Hopf bifurcation of the whirl is the value of $\Delta$ at which the complex pair of eigenvalues of the Jacobian with the largest real part crosses the imaginary axis. The period doubling and halving of the modulated whirl are the points where a Floquet multiplier of the periodic orbit passes through $-1$. We compute the multipliers from the Poincar\'e map on the section where $\mathrm{Re}\,V$ takes its value at the steady whirl, using H\'enon's trick to locate the crossings of the section~\cite{Henon1982}, \cite[App.~E]{Tufillaro1992}. For the two-mode string the Hopf
bifurcation is the Neimark--Sacker point of the periodic whirl, where a
complex pair of Floquet multipliers of the stroboscopic map crosses the unit
circle. The Jacobian and Floquet points are found by bisection to $2\times10^{-5}$ in $\Delta$. For the
GE string the dimension is too high for a linearization, and we use instead the rate at
which a small modulation of the whirl grows or decays: a run of 150 units of
$\tau$ starting on the whirl with a modulation of 0.002 added, the envelope of $\mathrm{Re}\,V$ about its running mean fitted by an exponential while it lies
between $10^{-3}$ and $5\times10^{-2}$, and the zero of the rate against
$\Delta$, at steps of 0.005, extrapolated from a line through the two decay
rates nearest it. (The decay rates are used because the growth rates past
the bifurcation are biased low by saturation of the envelope.) Applied to
the averaged and two-mode models, this growth-rate method reproduces the Jacobian and Floquet values to within $3\times10^{-4}$ in every case. We take $3\times10^{-4}$ as the uncertainty of the growth-rate values for the GE string.

Where no linearization is at hand, which is the period doubling of the
two-mode and GE strings, we use an order parameter~\cite{Landau1937}: a quantity that is zero on one side of the bifurcation and grows linearly in $\Delta$ on the other. We fit a line to it from the scan and take its zero as the bifurcation point. For the period doublings the order parameter is $d^2$, where $d$ is the median of $|\mathrm{Im}\,V_{n+1} - \mathrm{Im}\,V_n|$ and $V_n$ is the value of $V$ at the $n$th upward crossing of $\mathrm{Re}\,V$ through its mean; the line is fitted from the first of three consecutive points at which $d$ exceeds 0.02, over the following 0.012 in $\Delta$. For the Hopf bifurcation, where we use the scan extrapolation only as a check, the order parameter is the variance of $\mathrm{Re}\,V$. The
uncertainty we attach to such an estimate is half the spread of the zero
over fit windows of 0.008, 0.012 and 0.016 and thresholds scaled by 0.75
and 1.5. Against the exact values for the averaged equations the
extrapolated period doublings are off by $+0.0008$ ($1\to2$) and $+0.0015$
($2\to1$) at $\gamma = 0.45$ and by $-0.0017$ at $\gamma = 0.50$, which is
the size of the systematic error to allow for in the string values. The
extrapolated Hopf point is worse, by up to $+0.004$ at $\gamma = 0.50$,
because just past the Hopf bifurcation the modulation grows so slowly that
it has not saturated after the 300 units of transient; this is why the
growth-rate method is used instead. The first whirl-reversing attractor and
the first non-periodic attractor are located only to within the scan step,
and, as Sec.~\ref{sec:qual} shows, depend on which of several coexisting
attractors the run reaches.

The primary homoclinic orbit is found by following the unstable manifold of
the upper planar saddle focus. We start a run on the planar state with
$W = 10^{-5}$ along the unstable eigenvector, follow it until $|W|^2$
reaches its first local minimum after the excursion, and measure the signed
distance $\beta$ of that return from the stable manifold, as the projection onto
the left unstable eigenvector. The zero of $\beta(\Delta)$, interpolated
linearly between the bracketing points at steps of 0.005 ($\gamma = 0.45$)
or 0.01 ($\gamma = 0.50$), marks the homoclinic orbit; $\beta$ is the splitting function of
Sec.~\ref{sec:terms}, and the uncertainty we attach to $\Delta^*$ is the
largest change in the zero when either bracketing point is replaced by its
outer neighbour. For the GE and two-mode models the planar state is found by running the string with the whirl suppressed for 60 units of $\tau$. Its in-plane spiral rate and frequency are measured by starting two runs from the settled state, one of them with a small in-plane kick (0.005 in $V$), and fitting a linear map to the difference of the two, in which the common transient and the drive-frequency ripple cancel; $\lambda_u$ is fitted from the growth of the seed.

Because the GE and two-mode strings are periodically forced, their planar state is a periodic orbit, and a
homoclinic orbit to it is, generically, a transverse intersection of its
stable and unstable manifolds, which persists over an interval of $\Delta$
rather than occurring at an isolated value. The averaging theorem, which
controls the slow flow over times of order $1/\eps$, says nothing about such
global objects; the manifolds that coincide at $\Delta^*$ in the averaged
flow may split by an amount beyond all orders in $\eps$ in the full
system~\cite{Guckenheimer1983,Neishtadt1984,Holmes1988}. What we locate in
the GE and two-mode models is therefore the zero of a slow-flow proxy, the demodulated
and period-averaged distance $\beta$ of the first return from the plane, measured
in the same way as in the averaged equations. Its zero marks where the first
return of the string's unstable manifold passes through the plane in the
averaged sense, and it is this quantity, not an isolated homoclinic orbit of
the string, that the comparison between the models is about.

Two checks against known results support the methods. For the averaged
equations the Hopf bifurcation found by continuation of the whirl branch
($\Delta = 2.6863$ at $\gamma = 0.45$) agrees with the Jacobian, and the
homoclinic values $\Delta^* = 3.0783$ and 3.4032 agree with O'Reilly's 3.078
and 3.405~\cite{OReilly1993}. In the averaged equations the in-plane decay rate is exactly $\gamma$; our measurement returns it to within 0.6\%, and $\omega$ and $\lambda_u$ to within 0.5\%. The same measurement returns $|\rho| = \gamma$ to within 1\% in the two-mode equations and 0.3\% in the GE string.

% ============================================================================
\section{Qualitative agreement of the three models}
\label{sec:qual}

\begin{figure}[t]
\centering
\includegraphics[width=\textwidth]{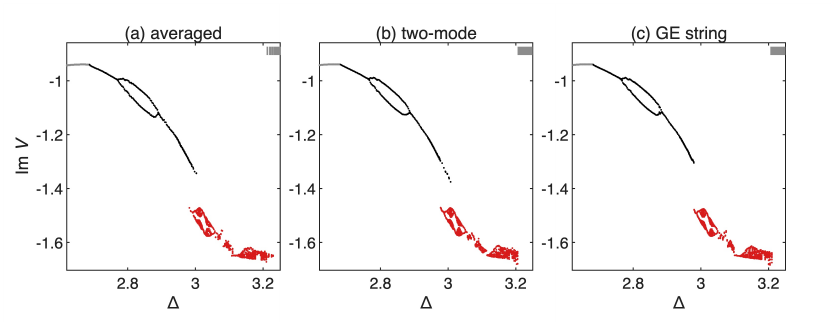}
\caption{Bifurcation diagrams at $\gamma = 0.45$: the value of $\mathrm{Im}\,V$ at each upward crossing of $\mathrm{Re}\,V$
through its mean, plotted against the scaled
detuning $\Delta$. (a) The averaged equations; (b) the two-mode equations and (c) the GE string, both at $\sigma = 1$~s$^{-1}$ ($\eps = 0.0051$). Gray: steady whirl. Black: whirl that keeps
its direction. Red: whirl that reverses its direction. Gray ticks along
the top edge mark runs that have fallen to planar motion, past the crisis.
GE and two-mode values of $\Delta$ are corrected for the frequency offsets of their integrators (Sec.~\ref{sec:methods}).}
\label{fig:bif045}
\end{figure}

Figure~\ref{fig:bif045} shows brute-force bifurcation diagrams for
$\gamma = 0.45$. At the left of each panel the string whirls steadily (gray).
Near $\Delta = 2.68$ the steady whirl loses stability in a Hopf bifurcation,
and the amplitude of the whirl begins to oscillate slowly (black). Near 2.77
this modulation doubles its period, and near 2.89 it halves again, closing a period bubble. Between 2.97 and 3.01, depending on the model, a
new kind of attractor appears (red), on which the angular momentum changes
sign and the whirl reverses its direction. These reversing attractors double
their period and become chaotic within about 0.03 in $\Delta$, and the
chaotic region continues past the primary homoclinic orbit at $\Delta^*
\approx 3.08$, ending in a crisis near $\Delta = 3.2$. The sequence is the same in the three panels, and in the string computations at $\sigma = 2$ and 4~s$^{-1}$, which are not shown; on this scale the diagrams nearly lie on top of one another.

\begin{figure}[t]
\centering
\includegraphics[width=\textwidth]{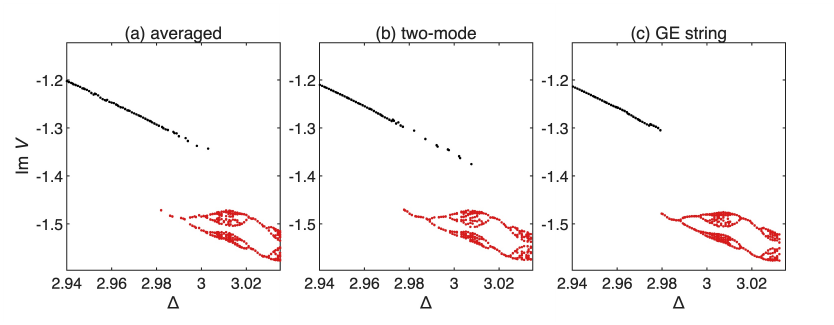}
\caption{The region of Fig.~\ref{fig:bif045} where whirl-reversing
attractors appear, from a scan with step 0.001 in $\Delta$: (a) the averaged equations; (b) the two-mode equations and (c) the GE string, both at $\sigma = 1$~s$^{-1}$.}
\label{fig:zoom}
\end{figure}

Figure~\ref{fig:zoom} enlarges the region where the reversing attractors
appear. In every model the reversing attractor coexists over part of this
range with the whirl that keeps its direction, and a run that starts from
the steady-whirl state may land on either one. For example, the two-mode
string at $\sigma = 1$~s$^{-1}$ first reverses at $\Delta = 2.9775$, but
in the scan with step 0.001 it lands again on the whirl that keeps its direction at 2.982, 2.987, 2.992--2.9925, 2.997--2.9975, 3.002--3.0025 and 3.0075 (nominal scan values; the correction of Sec.~\ref{sec:methods} is 0.0002 here),
with reversing attractors on either side of each. Where the black branch ends in a scan therefore
reflects the basins of the two attractors as well as their bifurcations,
and it does not move monotonically with $\eps$. We return to the coexisting attractors in
Sec.~\ref{sec:steps}.

At $\gamma = 0.50$ the route, in diagrams not shown here, begins in the same way,
with a Hopf bifurcation near $\Delta = 2.65$ and a period doubling near 2.92,
but the period-two whirl does not halve again. Instead the scans jump to
reversing attractors near $\Delta = 3.05$--3.11, which double and become
chaotic near 3.13--3.17. The primary homoclinic orbit lies further out, at
$\Delta^* = 3.26$--3.40, and since $\gamma/\lambda_u > 1$ here it does not
itself produce chaos. Coexisting attractors are again present; in the
averaged equations, for instance, period-two whirls and reversing orbits
alternate along the scan between $\Delta = 3.11$ and 3.16.

\begin{figure}[t]
\centering
\includegraphics[width=\textwidth]{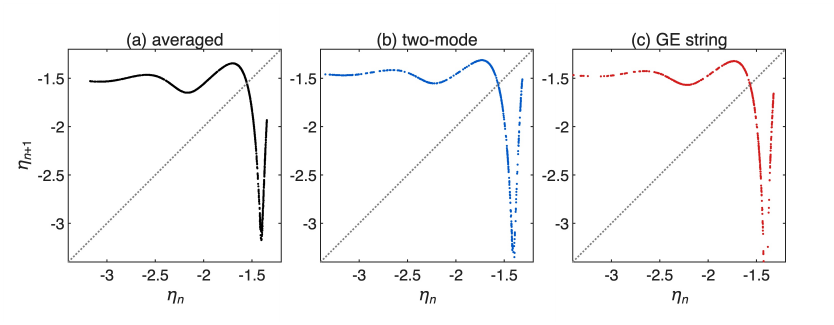}
\caption{First-return maps at $\gamma = 0.45$, $\Delta = 3.17$: the
value of $\eta = \log_{10}|W|^2$ at successive minima of $|W|^2$.
(a) averaged equations, 1929 returns; (b) two-mode string, 616 returns;
(c) GE string, 631 returns; both at $\sigma = 2$~s$^{-1}$
($\eps = 0.0103$).}
\label{fig:maps}
\end{figure}

The chaotic attractors themselves are also alike. Figure~\ref{fig:maps}
shows first-return maps at $\gamma = 0.45$, $\Delta = 3.17$, taken at the
successive minima of the out-of-plane energy $|W|^2$, which is where the
string comes closest to planar motion between whirling bursts. In all three
models the map has the same sequence of monotone branches, with a shallow
maximum, a minimum, a sharp maximum, and a steep descent into a deep
minimum, which belongs to passages close to the saddle focus. The critical
points of the branches agree to within about 0.05 in $\eta = \log_{10}|W|^2$.
The two string maps are nearly indistinguishable from each other; the
averaged map has the same shape but differs in the depth of the middle
minimum and the extent of the attractor, much as the averaged map itself
does at a slightly different $\Delta$.

The largest Lyapunov exponents come from sweeps in $\Delta$, up and down in steps of 0.01, in which each run
starts on the attractor of the previous one, discards 50 units of $\tau$,
and then follows two nearby trajectories for 150 units; the averaged exponent is the Benettin estimate from twin trajectories renormalized every unit of $\tau$, the two-mode exponent is the Benettin estimate with renormalization every drive period~\cite{Benettin1980}, and the GE exponent is the slope of the
logarithmic separation of twin runs, one of which has $w$ perturbed by
1~nm, fitted while the separation lies between $10^2$ times its initial
value and $10^{-2}$ of the amplitude. Measured per unit of slow time, the exponents are of the same size
in the three models, within a factor of two. Averaged over $3.13 \le \Delta \le 3.20$ at $\gamma = 0.45$,
$\lambda/c$ is $0.063 \pm 0.010$ for the averaged equations; 0.051, 0.069,
0.053 and 0.046 for the two-mode string at $\sigma = 0.5$, 1, 2 and
4~s$^{-1}$; and 0.034, 0.054, 0.052 and 0.061 for the GE string. The quoted
uncertainty is the standard error of the mean over the band; single values
scatter by about $\pm 0.03$. The upper crisis, where the chaotic attractor is
destroyed and the string falls to small-amplitude planar motion, is reached
by the scans of Fig.~\ref{fig:bif045} and by the sweeps used for the Lyapunov exponents. At
$\sigma = 1$~s$^{-1}$ the first planar run occurs at $\Delta = 3.2125$ in
the averaged equations and at 3.21 in the GE and two-mode models; at
$\sigma = 4$~s$^{-1}$ the two-mode string falls to planar motion from
3.1925. Near the crisis the transients are long, and non-planar runs recur
among the planar ones (up to $\Delta = 3.23$ in the averaged equations), so
the first planar run locates the crisis only to about $\pm 0.01$; the crisis points are
the least certain of the bifurcation points. After the crisis all three models settle
on the same planar state.

Finally, the Shil'nikov ratio $|\rho|/\lambda_u$ at the primary homoclinic
orbit, which decides whether the orbit is accompanied by horseshoes, is the
same in all three models to within 1.5\% for $\eps \le 0.01$ and 2.4\% at $\eps \approx 0.02$ (Table~\ref{tab:sf}). Because the in-plane pair has real part $-\gamma$ in every model, the agreement of the ratio is that of the unstable rate $\lambda_u$.
The Shil'nikov ratio is about 0.9 at $\gamma = 0.45$ and about 1.24 at $\gamma = 0.50$, so all three models agree that at $\gamma = 0.45$ the homoclinic orbit is accompanied by horseshoes and at $\gamma = 0.50$ it is not; the chaos found at $\gamma = 0.50$ arises away from the homoclinic orbit.

\begin{table}[t]
\centering
\caption{The upper planar saddle focus at the primary homoclinic orbit:
in-plane decay rate $|\rho|$, rotation frequency $\omega$, unstable rate
$\lambda_u$, and the Shil'nikov ratio $|\rho|/\lambda_u$, in units of the
slow time. Each entry is the median over the six runs nearest $\Delta^*$,
with half the range over those runs in parentheses, in units of the last
digit. $|\rho|$ and $\omega$ come from the response of the settled planar state to a small in-plane kick (Sec.~\ref{sec:protocol}); $\lambda_u$ comes from the growth of the out-of-plane seed. Fits that fail outright (more than 5\% from the median of the six, which happens in one or two runs of six in some rows) are excluded. The ratio is computed from the unrounded medians. For the averaged equations the exact values
from the eigenvalues are 0.450, 1.884, 0.493 and 0.912 at $\gamma = 0.45$,
and 0.500, 1.744, 0.405 and 1.235 at $\gamma = 0.50$; the measured values are
listed so that the comparison is between like quantities.}
\label{tab:sf}
\begin{tabular}{llcccc}
\toprule
$\gamma$ & model & $|\rho|$ & $\omega$ & $\lambda_u$ & $|\rho|/\lambda_u$ \\
\midrule
0.45 & averaged & 0.453(0) & 1.878(0) & 0.493(1) & 0.917 \\
 & two-mode, $\sigma = 1$ & 0.451(0) & 1.866(0) & 0.496(1) & 0.909 \\
  & two-mode, $\sigma = 2$ & 0.451(0) & 1.848(0) & 0.498(1) & 0.905 \\
 & two-mode, $\sigma = 4$ & 0.450(0) & 1.818(0) & 0.502(1) & 0.896 \\
  & GE, $\sigma = 1$ & 0.450(2) & 1.874(5) & 0.495(1) & 0.910 \\
 & GE, $\sigma = 2$ & 0.451(2) & 1.840(4) & 0.497(1) & 0.907 \\
  & GE, $\sigma = 4$ & 0.450(2) & 1.809(4) & 0.500(1) & 0.900 \\
\midrule
0.50 & averaged & 0.502(0) & 1.736(2) & 0.405(0) & 1.240 \\
  & two-mode, $\sigma = 1$ & 0.501(0) & 1.732(2) & 0.407(0) & 1.230 \\
 & two-mode, $\sigma = 2$ & 0.500(0) & 1.719(2) & 0.409(0) & 1.222 \\
  & two-mode, $\sigma = 4$ & 0.500(0) & 1.698(2) & 0.413(1) & 1.210 \\
  & GE, $\sigma = 1$ & 0.501(2) & 1.723(12) & 0.406(0) & 1.233 \\
  & GE, $\sigma = 2$ & 0.501(1) & 1.714(13) & 0.408(0) & 1.229 \\
  & GE, $\sigma = 4$ & 0.500(2) & 1.690(7) & 0.411(1) & 1.218 \\
\bottomrule
\end{tabular}
\end{table}

% ============================================================================
\section{Quantitative comparison of the three models}
\label{sec:steps}

\begin{figure}[!t]
\centering
\includegraphics[width=0.9\textwidth]{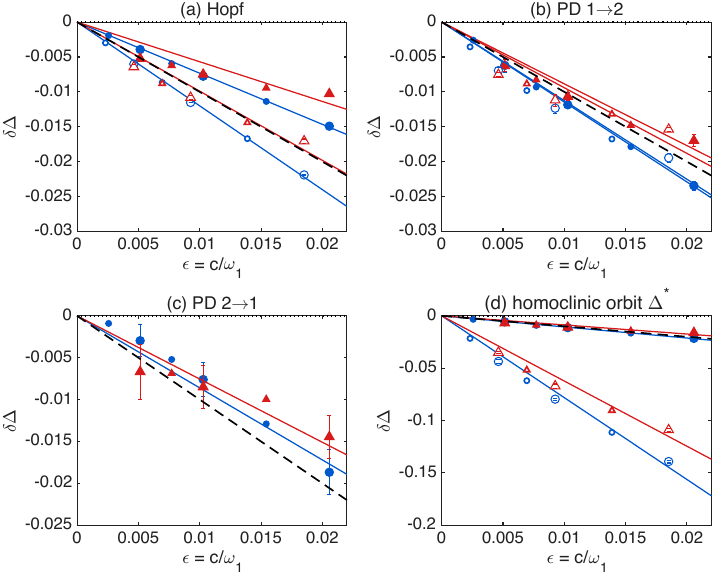}
\caption{Shift of each bifurcation point from its value in the averaged equations,
$\delta\Delta = \Delta_{\mathrm{model}} - \Delta_{\mathrm{avg}}$, against
$\eps = c/\omega_1$, with the uncertainties of Tables~\ref{tab:g045}
and~\ref{tab:g050} as error bars where they are larger than the symbols.
Circles (blue), two-mode string; triangles (red), GE string; filled
symbols, $\gamma = 0.45$; open symbols, $\gamma = 0.50$. Small symbols, the two-mode string at $\sigma = 0.5$, 1.5 and 3~s$^{-1}$ and the GE string at $\sigma = 1.5$ and 3~s$^{-1}$, which are not used in the fits ($\sigma = 6$~s$^{-1}$ lies beyond the plotted range). Lines are the fits through the origin; the dashed line is $\delta\Delta = -\eps$.}
\label{fig:shifts}
\end{figure}

Tables~\ref{tab:g045} and~\ref{tab:g050} list the locations of the
bifurcation points in every model, with the method by which each was found and its
uncertainty, and Fig.~\ref{fig:shifts} plots the shifts
$\delta\Delta = \Delta_{\mathrm{model}} - \Delta_{\mathrm{avg}}$ against
$\eps$. The last rows of each table give the slope of a least-squares fit of the shift against $\eps$ over the three values of $\sigma$, through the origin, because every shift must vanish as $\eps \to 0$. 

\begin{table}[t]
\centering
\caption{Bifurcation points at $\gamma = 0.45$: the value of $\Delta$ at the Hopf
(Neimark--Sacker) bifurcation of the steady whirl, the period doubling
$1\to2$, the period halving $2\to1$ that closes the bubble, and the primary
homoclinic orbit $\Delta^*$. Superscripts give the method
(Sec.~\ref{sec:protocol}): J, Jacobian of the averaged equilibrium; F,
Floquet multipliers; G, zero of the growth rate of a small modulation (GE string); S, extrapolation of the scan; H, zero of the splitting
function. Parentheses give the uncertainty in units of the last digit where
it exceeds that digit; brackets give the shift from the averaged value. The
last rows give the slope $\delta\Delta/\eps$ of a least-squares fit of the shift through the origin.}
\label{tab:g045}
\small
\resizebox{\textwidth}{!}{\begin{tabular}{lccccc}
\toprule
model & $\eps$ & Hopf & PD $1\to2$ & PD $2\to1$ & $\Delta^*$ \\
\midrule
averaged & --- & 2.6863$^{\mathrm{J}}$ & 2.7722$^{\mathrm{F}}$ & 2.8888$^{\mathrm{F}}$ & 3.0783(14)$^{\mathrm{H}}$ \\
two-mode, $\sigma = 1$ & 0.0051 & 2.6823$^{\mathrm{F}}$ [$-0.0039$] & 2.7660(10)$^{\mathrm{S}}$ [$-0.0062$] & 2.8859(19)$^{\mathrm{S}}$ [$-0.0030$] & 3.0726(9)$^{\mathrm{H}}$ [$-0.0058$] \\
two-mode, $\sigma = 2$ & 0.0103 & 2.6785$^{\mathrm{F}}$ [$-0.0078$] & 2.7603(3)$^{\mathrm{S}}$ [$-0.0119$] & 2.8812(20)$^{\mathrm{S}}$ [$-0.0076$] & 3.0670(7)$^{\mathrm{H}}$ [$-0.0113$] \\
two-mode, $\sigma = 4$ & 0.0206 & 2.6714$^{\mathrm{F}}$ [$-0.0149$] & 2.7487(6)$^{\mathrm{S}}$ [$-0.0234$] & 2.8702(27)$^{\mathrm{S}}$ [$-0.0187$] & 3.0568(6)$^{\mathrm{H}}$ [$-0.0215$] \\
GE, $\sigma = 1$ & 0.0051 & 2.6811(3)$^{\mathrm{G}}$ [$-0.0052$] & 2.7659(8)$^{\mathrm{S}}$ [$-0.0063$] & 2.8822(33)$^{\mathrm{S}}$ [$-0.0067$] & 3.0714(16)$^{\mathrm{H}}$ [$-0.0070$] \\
GE, $\sigma = 2$ & 0.0103 & 2.6788(3)$^{\mathrm{G}}$ [$-0.0075$] & 2.7615(5)$^{\mathrm{S}}$ [$-0.0107$] & 2.8803(25)$^{\mathrm{S}}$ [$-0.0085$] & 3.0672(15)$^{\mathrm{H}}$ [$-0.0111$] \\
GE, $\sigma = 4$ & 0.0206 & 2.6760(3)$^{\mathrm{G}}$ [$-0.0103$] & 2.7552(9)$^{\mathrm{S}}$ [$-0.0170$] & 2.8744(25)$^{\mathrm{S}}$ [$-0.0145$] & 3.0621(11)$^{\mathrm{H}}$ [$-0.0162$] \\
\midrule
\multicolumn{2}{l}{slope, two-mode} & $-0.73$ & $-1.15$ & $-0.86$ & $-1.06$ \\
\multicolumn{2}{l}{slope, GE} & $-0.57$ & $-0.89$ & $-0.75$ & $-0.87$ \\
\bottomrule
\end{tabular}}
\end{table}

\begin{table}[t]
\centering
\caption{Bifurcation points at $\gamma = 0.50$, as in Table~\ref{tab:g045}. The
period-two whirl does not halve at this damping.}
\label{tab:g050}
\small
\resizebox{\textwidth}{!}{\begin{tabular}{lcccc}
\toprule
model & $\eps$ & Hopf & PD $1\to2$ & $\Delta^*$ \\
\midrule
averaged & --- & 2.6500$^{\mathrm{J}}$ & 2.9256$^{\mathrm{F}}$ & 3.4032(9)$^{\mathrm{H}}$ \\
two-mode, $\sigma = 1$ & 0.0046 & 2.6440$^{\mathrm{F}}$ [$-0.0060$] & 2.9186(4)$^{\mathrm{S}}$ [$-0.0070$] & 3.3596(17)$^{\mathrm{H}}$ [$-0.0436$] \\
two-mode, $\sigma = 2$ & 0.0093 & 2.6385$^{\mathrm{F}}$ [$-0.0115$] & 2.9133(8)$^{\mathrm{S}}$ [$-0.0123$] & 3.3236(11)$^{\mathrm{H}}$ [$-0.0796$] \\
two-mode, $\sigma = 4$ & 0.0185 & 2.6281$^{\mathrm{F}}$ [$-0.0219$] & 2.9062(6)$^{\mathrm{S}}$ [$-0.0195$] & 3.2640(14)$^{\mathrm{H}}$ [$-0.1392$] \\
GE, $\sigma = 1$ & 0.0046 & 2.6436(3)$^{\mathrm{G}}$ [$-0.0064$] & 2.9181(3)$^{\mathrm{S}}$ [$-0.0075$] & 3.3682(7)$^{\mathrm{H}}$ [$-0.0350$] \\
GE, $\sigma = 2$ & 0.0093 & 2.6392(3)$^{\mathrm{G}}$ [$-0.0108$] & 2.9144(9)$^{\mathrm{S}}$ [$-0.0112$] & 3.3361(20)$^{\mathrm{H}}$ [$-0.0671$] \\
GE, $\sigma = 4$ & 0.0185 & 2.6330(3)$^{\mathrm{G}}$ [$-0.0171$] & 2.9102(3)$^{\mathrm{S}}$ [$-0.0154$] & 3.2942(15)$^{\mathrm{H}}$ [$-0.1090$] \\
\midrule
\multicolumn{2}{l}{slope, two-mode} & $-1.20$ & $-1.13$ & $-7.82$ \\
\multicolumn{2}{l}{slope, GE} & $-0.99$ & $-0.94$ & $-6.23$ \\
\bottomrule
\end{tabular}}
\end{table}

\subsection{Averaging (step C)}
\label{sec:stepC}

In the two-mode string every bifurcation point moves to lower $\Delta$ by an amount proportional to $\eps$: for the precisely located points (the Hopf points, and $\Delta^*$ at $\gamma = 0.45$) a straight line through the three shifts meets $\eps = 0$ within 0.001 of the origin, and for the period doublings within 0.004, which is the size of the systematic error of the scan extrapolation (Sec.~\ref{sec:protocol}). The slopes through the origin lie between $-0.7$ and $-1.2$ for the local bifurcations and for $\Delta^*$ at $\gamma = 0.45$. Four more damping values for the two-mode string, $\sigma = 0.5$, 1.5, 3 and 6~s$^{-1}$ (small circles in Fig.~\ref{fig:shifts}), extend the test over a twelvefold range of $\eps$: $\delta\Delta/\eps$ for the Hopf point stays between $-0.70$ and $-0.77$ at $\gamma = 0.45$ and between $-1.13$ and $-1.30$ at $\gamma = 0.50$, and for $\Delta^*$ at $\gamma = 0.45$ between $-0.99$ and $-1.19$, with the value at $\sigma = 0.5$~s$^{-1}$, $-1.3 \pm 0.8$, within its uncertainty. This is the expected form of the averaging error: the
averaged equations place each of these bifurcations too high in drive
frequency by about $\eps$ in $\Delta$, which is $c\,\eps = c^2/\omega_1$ in
$\Omega$. At $\sigma = 1$~s$^{-1}$ this is about 2~mHz, far below any
practical frequency resolution; at $\sigma = 4$~s$^{-1}$ it is about
30~mHz, comparable to the width of a periodic window.

The periodic windows inside the chaotic band also move to lower $\Delta$ as $\eps$ grows. The first
period-one window at $\gamma = 0.45$ opens at $\Delta = 3.060$ in the
averaged equations, at 3.053, 3.048 and 3.038 in the two-mode string, and at
3.052, 3.046 and 3.042 in the GE string, for $\sigma = 1$, 2 and
4~s$^{-1}$.

The global events do not all behave like the local ones. The primary
homoclinic orbit at $\gamma = 0.45$ shifts like a local bifurcation, with a slope of $-1.06$ in the two-mode string. At $\gamma = 0.50$ it shifts five to
eight times as far, by $-0.035$ to $-0.14$ in $\Delta$, and here the shift is not proportional to $\eps$: in the two-mode string $\delta\Delta^*/\eps$ is $-9.3$ and $-9.4$ at $\sigma = 0.5$ and 1~s$^{-1}$ and falls to $-7.5$ at $\sigma = 4$~s$^{-1}$ and $-6.7$ at $\sigma = 6$~s$^{-1}$, because the shift is no longer small enough to be linear (Sec.~\ref{sec:hom}). The crisis at $\gamma = 0.45$ moves by $-0.0025$ at
$\sigma = 1$~s$^{-1}$, by $-0.0075$ at $\sigma = 2$~s$^{-1}$ and by $-0.02$
at $\sigma = 4$~s$^{-1}$ in the two-mode string, that is, roughly like a
local bifurcation, but with an uncertainty of $\pm 0.01$
(Sec.~\ref{sec:qual}). In the scan with step 0.001 the first reversing attractor at
$\gamma = 0.45$ lies between $\Delta = 2.971$ and 3.006 and the first
non-periodic attractor between 2.988 and 3.008, with no consistent
dependence on $\eps$; at $\gamma = 0.50$ they lie 0.01 to
0.06 below the averaged values. Since both are set partly by which of the
coexisting attractors a run happens to reach (Sec.~\ref{sec:qual}), we do
not fit a slope to them. Locating the underlying bifurcations would
require continuation of the periodic orbits, which we have not done.

The saddle focus that organizes the chaos changes little
(Table~\ref{tab:sf}). In the two-mode string its rotation frequency $\omega$ is lower than in the averaged equations by 0.6, 1.6 and 3.2\% at $\gamma = 0.45$ and by 0.3, 1.0 and 2.2\% at $\gamma = 0.50$, for $\eps \approx 0.005$, 0.01 and 0.02, and its unstable rate $\lambda_u$ is higher by 0.5, 1.0 and 1.9\% and by 0.5, 1.1 and 2.0\%, a relative change of about $\eps$. The Shil'nikov ratio changes by $-0.9$, $-1.3$ and $-2.3\%$ and by $-0.8$, $-1.4$ and $-2.4\%$, mostly through $\lambda_u$. 

\subsection{Kirchhoff--Carrier approximation and modal truncation (steps A and B)}
\label{sec:stepAB}

The GE string agrees with the two-mode string at the two smaller dampings
and departs from it at the largest. At $\sigma = 1$ and 2~s$^{-1}$ the
local bifurcation points of the GE and two-mode models differ by at most 0.002, and $\Delta^*$
at $\gamma = 0.45$ by at most 0.0012, which is within or close to the
uncertainties (the period halving at $\sigma = 1$, whose GE value is uncertain by 0.003, differs by 0.004). The only larger difference is
$\Delta^*$ at $\gamma = 0.50$, where the GE value lies 0.009 to 0.013 above
the two-mode value (and 0.030 above it at $\sigma = 4$~s$^{-1}$); Sec.~\ref{sec:hom} explains why this bifurcation point
magnifies a small difference between the models. At $\sigma = 4$~s$^{-1}$
the GE shifts are about 70 to 80\% of the two-mode shifts for every
bifurcation point, so for the local bifurcations the GE string sits 0.004 to 0.007 above the two-mode string. The GE shift from the averaged equations is therefore not proportional to $\eps$ over the three values of $\sigma$, and its slopes through the origin are 77 to 87\% of the two-mode ones. In every case the GE and two-mode models differ from each
other by less than either differs from the averaged equations, and at the
two smaller dampings most of the difference between the averaged equations
and the full string comes from the averaging.

Whether the difference at $\sigma = 4$~s$^{-1}$ is an effect of $\eps$ or
of amplitude cannot be decided within the one-parameter family of
Sec.~\ref{sec:params}, where $s^2 \propto \sigma$ and $\eps$ and amplitude grow
together. To separate $\eps$ from amplitude we
repeated the GE scans at $\gamma = 0.45$ with the stiffness $K$ scaled by
2 at $\sigma = 2$~s$^{-1}$ and by 4 at $\sigma = 4$~s$^{-1}$, which keeps
$\eps$ at its $\sigma = 2$ and 4 values but returns the amplitude scale $s$
to its $\sigma = 1$ value of 2.5~mm (the sample rate and the
longitudinal-mode cutoff of the scheme were scaled with $\sqrt{K}$ so that
the same two longitudinal modes are kept, and the scheme's cubic
coefficient was checked against the Kirchhoff--Carrier value to 0.5\% at
each stiffness). The bifurcation points of these runs, all extrapolated from the
scans in the same way, are compared with the ordinary GE and two-mode
values in Table~\ref{tab:amp}. At $\sigma = 2$~s$^{-1}$ the amplitude
makes no difference that the scan can resolve. At $\sigma = 4$~s$^{-1}$,
reducing the amplitude moves the GE Hopf shift from $-0.011$ to $-0.014$
and the period-doubling shift from $-0.016$ to $-0.020$, against the
two-mode values $-0.015$ and $-0.024$; that is, it closes one half to
three quarters of the gap between the GE and two-mode models, and leaves the GE
string within 0.001 to 0.003 of the two-mode string (the period halving,
whose change of $0.0005$ is within its uncertainty of $0.003$, does not
resolve the effect). The departure of the
GE string at the largest damping is therefore mostly an amplitude effect,
from the approximations that separate the Kirchhoff--Carrier equations
from the geometrically exact ones at a whirl amplitude of 5~mm on a 1~m
string, and not a breakdown of the first-order picture in $\eps$.

\begin{table}[t]
\centering
\caption{Amplitude against $\eps$ at $\gamma = 0.45$: bifurcation points of the GE
string with the stiffness scaled by 2 and 4 ($K{\times}2$, $K{\times}4$),
which have the $\eps$ of $\sigma = 2$ and 4~s$^{-1}$ but the whirl
amplitude of $\sigma = 1$~s$^{-1}$, compared with the ordinary GE and
two-mode scans. All entries are scan extrapolations (method S) on the 0.005 grid of the stiffness-scaled scans, corrected for the frequency offset of each model's integrator (for the GE string, at each stiffness); brackets give the
shift from the averaged value. The two-mode rows are extrapolated on the same 0.005 grid, which is why they differ in the last digit from Table~\ref{tab:g045}.}
\label{tab:amp}
\small
\resizebox{\textwidth}{!}{\begin{tabular}{lcccccc}
\toprule
model & $\sigma$ & $\eps$ & $s/s_{\sigma=1}$ & Hopf & PD $1\to2$ & PD $2\to1$ \\
\midrule
GE & 1 & 0.0051 & 1.00 & 2.6814 [$-0.0048$] & 2.7659 [$-0.0063$] & 2.8808 [$-0.0080$] \\
GE & 2 & 0.0103 & 1.41 & 2.6786 [$-0.0077$] & 2.7614 [$-0.0108$] & 2.8813 [$-0.0075$] \\
GE, $K{\times}2$ & 2 & 0.0103 & 1.00 & 2.6780 [$-0.0083$] & 2.7609 [$-0.0113$] & 2.8814 [$-0.0075$] \\
two-mode & 2 & 0.0103 & 1.41 & 2.6788 [$-0.0075$] & 2.7602 [$-0.0119$] & 2.8812 [$-0.0077$] \\
\midrule
GE & 4 & 0.0206 & 2.00 & 2.6752 [$-0.0111$] & 2.7557 [$-0.0164$] & 2.8742 [$-0.0146$] \\
GE, $K{\times}4$ & 4 & 0.0206 & 1.00 & 2.6724 [$-0.0138$] & 2.7519 [$-0.0203$] & 2.8747 [$-0.0141$] \\
two-mode & 4 & 0.0206 & 2.00 & 2.6714 [$-0.0149$] & 2.7486 [$-0.0235$] & 2.8701 [$-0.0188$] \\
\bottomrule
\end{tabular}}
\end{table}

The GE and two-mode strings also agree on the saddle focus. At every damping the GE string gives the same $\omega$ and $\lambda_u$ as the two-mode string to within 0.6\%, and the same Shil'nikov ratio to within 0.7\%. One of the six GE kick runs at $\sigma = 4$~s$^{-1}$, $\gamma = 0.45$ failed to fit and is excluded, as the caption of Table~\ref{tab:sf} says.

The GE and two-mode models, then, agree to the resolution of the comparison at
amplitude scales $s$ up to 3.5~mm, and part company by 0.004 to 0.007 in $\Delta$ at 5~mm. GE Hopf points at $\sigma = 1.5$, 3 and 6~s$^{-1}$ show how the departure sets in: at both values of $\gamma$ the GE value lies within 0.0015 of the two-mode value up to $\sigma = 2$~s$^{-1}$ ($s = 3.5$~mm), and 0.002, 0.005 and 0.012 above it at $\sigma = 3$, 4 and 6~s$^{-1}$ ($s = 4.3$, 5.0 and 6.1~mm). Doubling $s^2$ from $\sigma = 3$ to 6~s$^{-1}$ multiplies the difference by about six. At $\sigma = 6$~s$^{-1}$ the GE string no longer follows the route: its modulated whirl is irregular and shows no period bubble, so its other bifurcation points cannot be located there; we have not checked whether this is a property of the string or of the scheme at this amplitude. We do not use the GE runs at $\sigma = 0.5$~s$^{-1}$: there the correction for the scheme's frequency offset, 0.0054 in $\Delta$, is larger than the shift being measured, so the points are only as good as the calibration of that offset; their Hopf and period-doubling points lie 0.0015 to 0.002 below the two-mode values. Our runs cannot separate step A from step B, but there
is reason to expect step B to be the smaller. In the Kirchhoff--Carrier
equation the tension is uniform along the string, so a single sine mode in
each polarization is an exact shape for the nonlinear term, and the higher
modes are excited only through the drive point, far from their resonances;
the GE string reproduces the two-mode cubic coefficient to 0.1\% at small
amplitude (Sec.~\ref{sec:methods}). Two effects that increase with the amplitude remain, and both belong to step A: the difference between the exact stretch and its quadratic approximation, and the inertia of the longitudinal motion, which the Kirchhoff--Carrier approximation neglects. The argument for step B rests on the symmetry of the string and
of a purely harmonic drive. A static load added to the drive breaks that
symmetry, and the second in-plane mode then couples to the first
out-of-plane mode in a 2:1 resonance~\cite{Bayly1998}, which a model with
one mode in each plane cannot represent. A string with such a load, a
stronger nonlinearity, an internal resonance with a higher mode, or a drive
near a longitudinal frequency would test both steps more
severely~\cite{Bayly1998,Leamy2000,Watzky1992}.

% ============================================================================
\section{Why the homoclinic bifurcation is more sensitive}
\label{sec:hom}

\begin{figure}[!t]
\centering
\includegraphics[width=\textwidth]{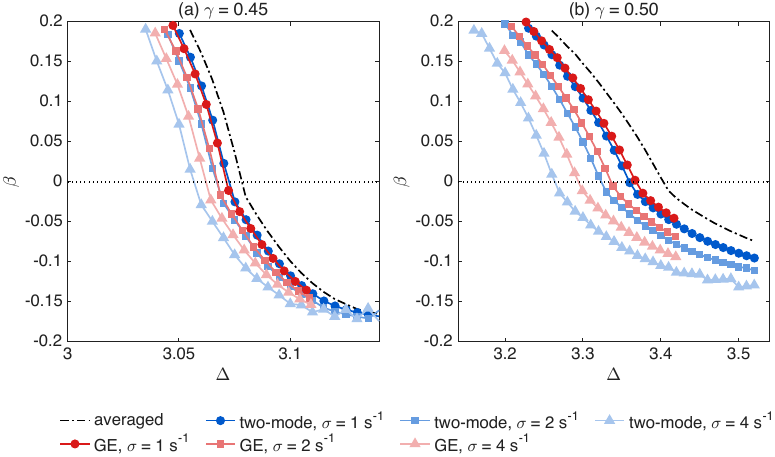}
\caption{Splitting function $\beta(\Delta)$ of the primary homoclinic loop: the
signed distance of the first return of the unstable manifold from the stable
manifold of the planar saddle focus. Its zero marks the homoclinic orbit.
Black, averaged equations; blue, two-mode string; red, GE string; markers
distinguish $\sigma = 1$, 2 and 4~s$^{-1}$. At $\gamma = 0.45$ the curves
become irregular further beyond the first zero (outside the plotted range), where the unstable manifold makes
several near returns; only the first zero is used.}
\label{fig:split}
\end{figure}

The difference between the homoclinic shifts at $\gamma = 0.45$ and $\gamma = 0.50$ is not a difference in the
size of the modeling error. Figure~\ref{fig:split} shows the splitting
function $\beta(\Delta)$ for all seven computations at both values of $\gamma$.
At the averaged $\Delta^*$, the two-mode string at $\sigma = 1$~s$^{-1}$ has
$\beta = -0.040$ at $\gamma = 0.45$ and $\beta = -0.043$ at $\gamma = 0.50$ (the GE
string, $-0.042$ and $-0.036$), so the string moves the return of the
homoclinic loop by about the same amount in both cases. At both values of
$\gamma$ the change $\delta \beta$ grows with $\eps$, though less than in
proportion: $-0.040$, $-0.061$, $-0.10$ and $-0.043$, $-0.069$, $-0.11$ for
$\sigma = 1$, 2, 4~s$^{-1}$. What differs is the slope. Near the zero, the averaged
$\beta(\Delta)$ falls at $d\beta/d\Delta = -11$ (from the bracketing pair of points)
or $-8.9$ (from the four nearest points) at $\gamma = 0.45$, but only at
$-1.65$ or $-1.5$ at $\gamma = 0.50$, and at $\gamma = 0.50$ the curve
flattens further as $\Delta$ decreases. To first order the shift of the
homoclinic bifurcation is
\begin{equation}
\delta\Delta^* \approx -\frac{\delta \beta}{d\beta/d\Delta} ,
\label{eq:shift}
\end{equation}
where $\delta \beta$ is the change in $\beta$ produced by the model at fixed
$\Delta$. Equation~(\ref{eq:shift}) says that a global bifurcation inherits
the $O(\eps)$ error of the reduced model, amplified by the reciprocal of the
slope with which the splitting function crosses zero. With the bracketing
slopes and the two-mode values of $\delta \beta$ at $\sigma = 1$~s$^{-1}$, Eq.~(\ref{eq:shift}) predicts $\delta\Delta^* = -0.0035$ at $\gamma = 0.45$, against the measured $-0.0058$, and $-0.026$ at $\gamma = 0.50$, against the measured $-0.044$.
Both estimates are small by a factor of 1.6 to 1.7, for the same reason.
In every model $\beta(\Delta)$ is steepest just before its zero and flattens
beyond it (Fig.~\ref{fig:split}), and since the string's zero lies below
the averaged one, the stretch of curve that connects the two zeros is the
flatter part of the string's curve, beyond its own zero. The mean slope between the two zeros is
$-6.8$ at $\gamma = 0.45$ and $-1.0$ at $\gamma = 0.50$, rather than $-11$
and $-1.65$; with these slopes Eq.~(\ref{eq:shift}) is exact by
construction. The point of the estimate is the ratio of the shifts at $\gamma = 0.50$ and $\gamma = 0.45$, and
that is accounted for: the predicted ratio is 7.3 and the measured ratio is 7.6. At the larger dampings the predictions fall short by factors that grow
from 1.6 at $\sigma = 1$~s$^{-1}$ to 2.3 at $\sigma = 4$~s$^{-1}$, as the
shift grows and the linearization worsens, while the ratio of the shifts
at $\gamma = 0.45$ and 0.50 stays between 6.5 and 7.6.

The slope of the splitting function is a property of the averaged flow, so
it can be computed before any comparison is made. A reduced model that
places a homoclinic bifurcation where the splitting function crosses zero with a small slope should be expected to shift the homoclinic bifurcation by many times the model's nominal error, even when the model places the local bifurcations to within that error. The same
reasoning applies to the crisis and to the other global events, though we
have not measured their splitting functions. It applies equally to the
difference between the GE and two-mode models: at $\gamma = 0.50$ their
splitting functions differ by 0.015 to 0.025 at fixed $\Delta$ for $\sigma = 1$ and 2~s$^{-1}$, and the slope of about $-2$ turns this into the difference of 0.009 to 0.013 in
$\Delta^*$ noted in Sec.~\ref{sec:steps}. At $\gamma = 0.45$, where
$\gamma/\lambda_u < 1$, the curve $\beta(\Delta)$ is irregular beyond its first
zero, as the unstable manifold begins to make several near returns to the
saddle; this is the parameter region where subsidiary homoclinic orbits and
the associated windows are expected~\cite{Glendinning1984,Gaspard1984}.

% ============================================================================
\section{Discussion}
\label{sec:disc}

The averaged equations agree with the two string models in the sequence of bifurcations on the route to chaos and in the Shil'nikov ratio of the saddle focus that organizes it. They also agree in the strength of the chaos, measured by the largest Lyapunov exponent, to within a factor of two, and in the shape of the chaotic attractor, as seen in its return map. They place the bifurcations with an error that is first order in $\eps = c/\omega_1$, the ratio of the modulation rate to the linear fundamental frequency. For local bifurcations the
coefficient of $\eps$ is of order one. For global bifurcations it can be
larger, by the factor $1/|d\beta/d\Delta|$, and this factor can be computed from the averaged flow. At the two smaller dampings the full string adds nothing
beyond the two-mode model that our uncertainties can resolve, except at the homoclinic point of Sec.~\ref{sec:hom}, so there the
first-order error of averaging is the error that matters; at the largest
damping the full string's shifts are smaller than the two-mode string's by
20 to 30\%, mostly because of the larger amplitude there (Sec.~\ref{sec:stepAB}).

The invariants of the saddle focus also suggest a practical test of a reduced model. When the damping is the same on every in-plane degree of freedom, as here, the in-plane pair has real part $-\gamma$ and the Shil'nikov ratio is $\gamma/\lambda_u$, so the test needs only the unstable rate $\lambda_u$, found from one escape from the planar state; it can be computed in a high-dimensional model at a small fraction of the cost of a bifurcation scan. Here $\lambda_u$ agrees among the three models to within about 1\% for $\eps \le 0.01$ and 2\% at $\eps \approx 0.02$, even where the homoclinic orbit it belongs to has moved by several times $\eps$. Agreement of the Shil'nikov ratio is evidence that a reduction preserves the kind of chaos it predicts, whether or not it places that chaos at the right drive frequency.

The natural next step is a theory of the $O(\eps)$ shifts. Second-order
averaging of Eqs.~(\ref{eq:twomode}) would give a corrected slow flow in
which the Hopf bifurcation, the period doublings and the splitting function
could be computed directly~\cite{Sanders2007}.

The theory of spectral submanifolds offers a more general route~\cite{Haller2016,Breunung2018}. In
the language of spectral submanifolds the two-mode equations are a Galerkin projection onto the
linear modes, whereas a spectral submanifold is the nonlinear continuation of
the same modal subspace, an invariant manifold of the full system on which
the reduced dynamics is exact. For a periodically forced system the manifold
is time-periodic, and in polar coordinates in a frame rotating with the
drive its reduced dynamics become equations for slowly varying amplitudes
and phases~\cite{Breunung2018,Li2022a}. When the forced modes are in
internal resonance the master subspace contains all of them; for the two
transverse polarizations of the string, which are in 1:1 resonance, that
means two complex modes and a four-dimensional reduced model, and the
bifurcations of such reduced models, including quasi-periodic responses,
have been followed in the same framework~\cite{Li2022a,Li2022b}. Truncated
at leading order in the forcing, the spectral-submanifold equations for the slow amplitudes and phases
are autonomous and play the same role as the averaged equations used here,
and their higher-order terms are a natural source of corrections of the size
we measure. We have not computed the spectral submanifold of the string, and we
have not compared its leading-order equations with Eqs.~(\ref{eq:avg}) term
by term. The present results give a set of target numbers against which such
a calculation could
be checked: for example, a slope of $-0.7$ to $-1.2$ for the Hopf
bifurcation of the two-mode string, and a shift of $-0.044$ for the
homoclinic orbit at $\gamma = 0.50$ and $\eps = 0.0046$.

For an experiment the practical consequence is small. A shift of $\eps$ in
$\Delta$ is a shift of $c^2/\omega_1$ in drive frequency, a few millihertz
for a lightly damped wire. An experimenter who locates the chaotic band from
the averaged equations will find it where predicted, and will find its
internal structure, its windows and its crises, slightly displaced toward
lower frequency, by an amount that grows as the square of the damping.
The comparison here is between the models at a fixed mapping of the
physical drive onto the scaled force $\alpha$; how well that mapping holds
for a real drive is a separate question. For a wire driven at one end,
O'Reilly and Holmes found that the averaged equations predicted the
frequencies well but the drive amplitude only to within an order of
magnitude~\cite{OReilly1992}.

Several limitations should be kept in mind. All the computations are for one wire, a steel string 1~m long at a tension of 40~N, driven at one point; we varied its damping, and in one test its stiffness, but not its length, tension or drive point. In the scaled variables the averaged equations are the same for any string, so the error of averaging depends on the string only through $\eps$, but the size of the amplitude effects of steps A and B depends on the wire. We used two values of the scaled damping, one on each side of the Shil'nikov condition. All runs start from the same state, the steady whirl of the
averaged equations, and in regions of coexisting attractors this choice
affects what a scan finds. The period doublings, and the Hopf bifurcation of the GE string, are
located by extrapolation from brute-force scans rather than by continuation,
with the uncertainties of Sec.~\ref{sec:protocol}. We have compared the chaotic attractors through their return maps and Lyapunov exponents, and not through a template analysis. A template analysis of an experimental string, in the chaos that follows the torus-doubling cascade, found a horseshoe and a unimodal return map~\cite{Tufillaro1995}; near the Shil'nikov orbit studied here the return maps (Fig.~\ref{fig:maps}) are not unimodal, and the template of a chaos organized by a Shil'nikov orbit is not of finite type.

The extension that interests us most is to strings with an internal resonance. The whirl itself rests on one: the two polarizations of the fundamental have the same frequency, and the 1:1 resonance between them is what lets the planar motion lose its stability~\cite{Miles1965,Miles1984,OReilly1993}. Other resonances are easy to bring in. A static load breaks the symmetry of the string and couples its second in-plane mode to the first out-of-plane mode in a 2:1 resonance~\cite{Bayly1998}; the longitudinal motion enters the whirl through internal resonances~\cite{Leamy2000}; and unequal frequencies or unequal damping in the two polarizations, as on real supports, break the symmetry that the averaged equations assume. In each of these the motion departs from the symmetric one-mode whirl that the two-mode and averaged equations describe. For the resonances, a reduced model would need more than one mode in a plane, and the averaged equations more than two complex amplitudes. Whether analogues of the pattern found here hold in the more general case, with a few more degrees of freedom in play, is the question we would most like to answer next.

% ============================================================================
\section{Conclusions}
\label{sec:conc}

We have followed a complete route to chaos in three models of a driven
string: the geometrically exact equations with both polarizations and the
longitudinal motion, the two-mode Kirchhoff--Carrier equations, and Miles'
averaged equations. The three models show the same bifurcations in the same
order, the same coexistence of attractors, the same return maps, Lyapunov exponents of the same size, and the same verdict from the Shil'nikov
condition. Their bifurcations are shifted by amounts first order in the
ratio $\eps = c/\omega_1$ of the slow rate to the fundamental frequency. The shift of a local bifurcation is
about $\eps$ in the scaled detuning, and the shift of the homoclinic
bifurcation is larger where the splitting function of the homoclinic loop crosses zero with a small slope, by the reciprocal of its slope. The local bifurcations of the GE and
two-mode models agree with each other to within 0.002, or their uncertainties, in $\Delta$ at
$\eps \le 0.01$, so there the averaging, and not the truncation, accounts
for nearly all of the error; at $\eps = 0.02$ the full string's shifts are
70 to 80\% of the two-mode string's, and the difference is mostly one of
amplitude, not of $\eps$.

We offer these results as one worked example of what can be expected when a
reduced model is used near a route to chaos. The same comparison could be made for other distributed systems whose motion is confined to a few modes: a forced beam or plate, a pendulum on an elastic support, a fluid flow such as Taylor--Couette flow or Rayleigh--B\'enard convection~\cite{Curry1984}, or a laser with a few-mode cavity~\cite{Narducci1988}. It would be interesting to know whether the pattern found here, small and regular errors in the local bifurcations and larger ones in the global bifurcations set by the geometry of the homoclinic loop, holds more generally, and whether the guidance from the string example is accurate and useful there: to expect the local bifurcations within order $\eps$ of their averaged values, to compare the Shil'nikov ratio, and to estimate the shift of a global bifurcation from the slope of its splitting function.

% ============================================================================

\section*{CRediT authorship contribution statement}
\textbf{Nicholas B. Tufillaro:} Conceptualization, Methodology, Software,
Validation, Formal analysis, Investigation, Writing -- original draft,
Writing -- review and editing, Visualization.

\section*{Declaration of competing interest}
The author declares that he has no known competing financial interests or
personal relationships that could have appeared to influence the work
reported in this paper.

\section*{Funding}
This research did not receive any specific grant from funding agencies in
the public, commercial, or not-for-profit sectors.

\section*{Data availability}
The run data that support the findings of this study (the brute-force
scans, the growth-rate runs and the homoclinic-orbit runs for the three
models) and the MATLAB code for the three models, the scans and the
analysis are deposited at Zenodo~\cite{Tufillaro2026data}.

\section*{Declaration of generative AI and AI-assisted technologies in the manuscript preparation process}
During the preparation of this work the author used Claude (Anthropic) to document and test the computer code, structure the simulation sequences, and check the text of the paper. The author takes full responsibility for the content of the published article.

\bibliographystyle{elsarticle-num}
\bibliography{refs}

\begin{thebibliography}{10}
\expandafter\ifx\csname url\endcsname\relax
  \def\url#1{\texttt{#1}}\fi
\expandafter\ifx\csname urlprefix\endcsname\relax\def\urlprefix{URL }\fi
\expandafter\ifx\csname href\endcsname\relax
  \def\href#1#2{#2} \def\path#1{#1}\fi

\bibitem{Hanson1994}
R.~J. Hanson, J.~M. Anderson, H.~K. Macomber, Measurements of nonlinear effects
  in a driven vibrating wire, J. Acoust. Soc. Am. 96~(3) (1994) 1549--1556.
\newblock \href {https://doi.org/10.1121/1.410233}
  {\path{doi:10.1121/1.410233}}.

\bibitem{Molteno1990}
T.~C.~A. Molteno, N.~B. Tufillaro, Torus doubling and chaotic string
  vibrations: Experimental results, J. Sound Vib. 137~(2) (1990) 327--330.
\newblock \href {https://doi.org/10.1016/0022-460X(90)90796-3}
  {\path{doi:10.1016/0022-460X(90)90796-3}}.

\bibitem{Molteno2004}
T.~C. Molteno, N.~B. Tufillaro, An experimental investigation into the dynamics
  of a string, Am. J. Phys. 72~(9) (2004) 1157--1169.
\newblock \href {https://doi.org/10.1119/1.1764557}
  {\path{doi:10.1119/1.1764557}}.

\bibitem{Kirchhoff1876}
G.~R. Kirchhoff, Vorlesungen {\"u}ber mathematische {P}hysik, B. G. Teubner,
  Leipzig, 1883, three volumes.

\bibitem{Carrier1945}
G.~F. Carrier, On the non-linear vibration problem of the elastic string, Q.
  Appl. Math. 3~(2) (1945) 157--165.
\newblock \href {https://doi.org/10.1090/qam/12351}
  {\path{doi:10.1090/qam/12351}}.

\bibitem{Murthy1965}
G.~S.~S. Murthy, B.~S. Ramakrishna, Nonlinear character of resonance in
  stretched strings, J. Acoust. Soc. Am. 38~(3) (1965) 461--471.
\newblock \href {https://doi.org/10.1121/1.1909715}
  {\path{doi:10.1121/1.1909715}}.

\bibitem{Miles1965}
J.~W. Miles, Stability of forced oscillations of a vibrating string, J. Acoust.
  Soc. Am. 38~(5) (1965) 855--861.
\newblock \href {https://doi.org/10.1121/1.1909816}
  {\path{doi:10.1121/1.1909816}}.

\bibitem{Narasimha1968}
R.~Narasimha, Non-linear vibration of an elastic string, J. Sound Vib. 8~(1)
  (1968) 134--146.
\newblock \href {https://doi.org/10.1016/0022-460X(68)90200-9}
  {\path{doi:10.1016/0022-460X(68)90200-9}}.

\bibitem{Morse1968}
P.~M. Morse, K.~U. Ingard, Theoretical Acoustics, McGraw-Hill, New York, 1968.

\bibitem{Anand1969}
G.~V. Anand, Large-amplitude damped free vibration of a stretched string, J.
  Acoust. Soc. Am. 45~(5) (1969) 1089--1096.
\newblock \href {https://doi.org/10.1121/1.1911578}
  {\path{doi:10.1121/1.1911578}}.

\bibitem{Tagata1977}
G.~Tagata, Harmonically forced, finite amplitude vibration of a string, J.
  Sound Vib. 51~(4) (1977) 483--492.
\newblock \href {https://doi.org/10.1016/S0022-460X(77)80046-1}
  {\path{doi:10.1016/S0022-460X(77)80046-1}}.

\bibitem{Antman1980}
S.~S. Antman, The equations for large vibrations of strings, Am. Math. Mon.
  87~(5) (1980) 359--370.
\newblock \href {https://doi.org/10.1080/00029890.1980.11995034}
  {\path{doi:10.1080/00029890.1980.11995034}}.

\bibitem{Miles1984}
J.~W. Miles, Resonant, nonplanar motion of a stretched string, J. Acoust. Soc.
  Am. 75~(5) (1984) 1505--1510.
\newblock \href {https://doi.org/10.1121/1.390821}
  {\path{doi:10.1121/1.390821}}.

\bibitem{Johnson1989}
J.~M. Johnson, A.~K. Bajaj, Amplitude modulated and chaotic dynamics in
  resonant motion of strings, J. Sound Vib. 128~(1) (1989) 87--107.
\newblock \href {https://doi.org/10.1016/0022-460X(89)90682-2}
  {\path{doi:10.1016/0022-460X(89)90682-2}}.

\bibitem{Tufillaro1989}
N.~B. Tufillaro, Nonlinear and chaotic string vibrations, Am. J. Phys. 57~(5)
  (1989) 408--414.
\newblock \href {https://doi.org/10.1119/1.16011} {\path{doi:10.1119/1.16011}}.

\bibitem{OReilly1992}
O.~O'Reilly, P.~J. Holmes, Non-linear, non-planar and non-periodic vibrations
  of a string, J. Sound Vib. 153~(3) (1992) 413--435.
\newblock \href {https://doi.org/10.1016/0022-460X(92)90374-7}
  {\path{doi:10.1016/0022-460X(92)90374-7}}.

\bibitem{Mielke1992}
A.~Mielke, P.~Holmes, O.~O'Reilly, Cascades of homoclinic orbits to, and chaos
  near, a {H}amiltonian saddle-center, J. Dyn. Differ. Equ. 4~(1) (1992)
  95--126.
\newblock \href {https://doi.org/10.1007/BF01048157}
  {\path{doi:10.1007/BF01048157}}.

\bibitem{Bajaj1992}
A.~K. Bajaj, J.~M. Johnson, On the amplitude dynamics and crisis in resonant
  motion of stretched strings, Philos. Trans. R. Soc. Lond. A 338~(1649) (1992)
  1--41.
\newblock \href {https://doi.org/10.1098/rsta.1992.0001}
  {\path{doi:10.1098/rsta.1992.0001}}.

\bibitem{OReilly1993}
O.~O'Reilly, Global bifurcations in the forced vibration of a damped string,
  Int. J. Non-Linear Mech. 28~(3) (1993) 337--351.
\newblock \href {https://doi.org/10.1016/0020-7462(93)90040-R}
  {\path{doi:10.1016/0020-7462(93)90040-R}}.

\bibitem{Watzky1992}
A.~Watzky, Non-linear three-dimensional large-amplitude damped free vibration
  of a stiff elastic stretched string, J. Sound Vib. 153~(1) (1992) 125--142.
\newblock \href {https://doi.org/10.1016/0022-460X(92)90632-8}
  {\path{doi:10.1016/0022-460X(92)90632-8}}.

\bibitem{Gottlieb1996}
H.~P.~W. Gottlieb, Non-linear, non-planar transverse free vibrations of a
  constant-tension string, J. Sound Vib. 191~(4) (1996) 563--575.
\newblock \href {https://doi.org/10.1006/jsvi.1996.0140}
  {\path{doi:10.1006/jsvi.1996.0140}}.

\bibitem{Rubin1996}
M.~B. Rubin, O.~Gottlieb, Numerical solutions of forced vibration and whirling
  of a non-linear string using the theory of a {C}osserat point, J. Sound Vib.
  197~(1) (1996) 85--101.
\newblock \href {https://doi.org/10.1006/jsvi.1996.0518}
  {\path{doi:10.1006/jsvi.1996.0518}}.

\bibitem{Bayly1998}
P.~V. Bayly, K.~D. Murphy, Coupling between dissimilar modes in an
  asymmetrically forced string, J. Acoust. Soc. Am. 103~(6) (1998) 3362--3369.
\newblock \href {https://doi.org/10.1121/1.423050}
  {\path{doi:10.1121/1.423050}}.

\bibitem{Leamy2000}
M.~J. Leamy, O.~Gottlieb, Internal resonances in whirling strings involving
  longitudinal dynamics and material non-linearities, J. Sound Vib. 236~(4)
  (2000) 683--703.
\newblock \href {https://doi.org/10.1006/jsvi.2000.3039}
  {\path{doi:10.1006/jsvi.2000.3039}}.

\bibitem{Bilbao2005}
S.~Bilbao, Conservative numerical methods for nonlinear strings, J. Acoust.
  Soc. Am. 118~(5) (2005) 3316--3327.
\newblock \href {https://doi.org/10.1121/1.2046787}
  {\path{doi:10.1121/1.2046787}}.

\bibitem{Bilbao2006}
S.~Bilbao, Conservative numerical methods for nonlinear string dynamics:
  Non-planar vibration, Tech. rep., University of Edinburgh, arXiv:1905.09235
  (2006).

\bibitem{Ducceschi2022}
M.~Ducceschi, S.~Bilbao, Simulation of the geometrically exact nonlinear string
  via energy quadratisation, J. Sound Vib. 534 (2022) 117021.
\newblock \href {https://doi.org/10.1016/j.jsv.2022.117021}
  {\path{doi:10.1016/j.jsv.2022.117021}}.

\bibitem{Bilbao2023}
S.~Bilbao, M.~Ducceschi, Models of musical string vibration, Acoust. Sci.
  Technol. 44~(3) (2023) 194--209.
\newblock \href {https://doi.org/10.1250/ast.44.194}
  {\path{doi:10.1250/ast.44.194}}.

\bibitem{Russo2025}
R.~Russo, Non-iterative numerical simulation techniques for nonlinear string
  vibration in musical acoustics, Ph.D. thesis, University of Bologna (2025).

\bibitem{Suchde2016}
P.~Suchde, A.~S. Vasudeva~Murthy, On the wave equations of
  {K}irchhoff--{N}arasimha and {C}arrier, in: Applied Analysis in Biological
  and Physical Sciences, Springer Proceedings in Mathematics \& Statistics,
  Springer, New Delhi, 2016, pp. 259--275.
\newblock \href {https://doi.org/10.1007/978-81-322-3640-5_16}
  {\path{doi:10.1007/978-81-322-3640-5_16}}.

\bibitem{Suchde2017}
P.~Suchde, A.~S. Vasudeva~Murthy, The tangled history of the string equation,
  Bh\={a}van\={a} 1~(2),
  \url{https://bhavana.org.in/the-tangled-history-of-the-string-equation/}
  (2017).

\bibitem{Holmes1979}
P.~Holmes, A nonlinear oscillator with a strange attractor, Philos. Trans. R.
  Soc. Lond. A 292~(1394) (1979) 419--448.
\newblock \href {https://doi.org/10.1098/rsta.1979.0068}
  {\path{doi:10.1098/rsta.1979.0068}}.

\bibitem{Moon1979}
F.~C. Moon, P.~J. Holmes, A magnetoelastic strange attractor, J. Sound Vib.
  65~(2) (1979) 275--296.
\newblock \href {https://doi.org/10.1016/0022-460X(79)90520-0}
  {\path{doi:10.1016/0022-460X(79)90520-0}}.

\bibitem{Cusumano1995a}
J.~P. Cusumano, F.~C. Moon, Chaotic non-planar vibrations of the thin elastica,
  {P}art {I}: Experimental observation of planar instability, J. Sound Vib.
  179~(2) (1995) 185--208.
\newblock \href {https://doi.org/10.1006/jsvi.1995.0013}
  {\path{doi:10.1006/jsvi.1995.0013}}.

\bibitem{Cusumano1995b}
J.~P. Cusumano, F.~C. Moon, Chaotic non-planar vibrations of the thin elastica,
  {P}art {II}: Derivation and analysis of a low-dimensional model, J. Sound
  Vib. 179~(2) (1995) 209--226.
\newblock \href {https://doi.org/10.1006/jsvi.1995.0014}
  {\path{doi:10.1006/jsvi.1995.0014}}.

\bibitem{Pakdemirli1995}
M.~Pakdemirli, S.~A. Nayfeh, A.~H. Nayfeh, Analysis of one-to-one
  autoparametric resonances in cables---discretization vs. direct treatment,
  Nonlinear Dyn. 8~(1) (1995) 65--83.
\newblock \href {https://doi.org/10.1007/BF00045007}
  {\path{doi:10.1007/BF00045007}}.

\bibitem{Nayfeh1997}
A.~H. Nayfeh, W.~Lacarbonara, On the discretization of distributed-parameter
  systems with quadratic and cubic nonlinearities, Nonlinear Dyn. 13~(3) (1997)
  203--220.
\newblock \href {https://doi.org/10.1023/A:1008253901255}
  {\path{doi:10.1023/A:1008253901255}}.

\bibitem{Rega2004}
G.~Rega, Nonlinear vibrations of suspended cables---{P}art {I}: Modeling and
  analysis, Appl. Mech. Rev. 57~(6) (2004) 443--478.
\newblock \href {https://doi.org/10.1115/1.1777224}
  {\path{doi:10.1115/1.1777224}}.

\bibitem{Touze2006}
C.~Touz{\'e}, M.~Amabili, Nonlinear normal modes for damped geometrically
  nonlinear systems: Application to reduced-order modelling of harmonically
  forced structures, J. Sound Vib. 298~(4--5) (2006) 958--981.
\newblock \href {https://doi.org/10.1016/j.jsv.2006.06.032}
  {\path{doi:10.1016/j.jsv.2006.06.032}}.

\bibitem{Curry1984}
J.~H. Curry, J.~R. Herring, J.~Loncaric, S.~A. Orszag, Order and disorder in
  two- and three-dimensional {B}{\'e}nard convection, J. Fluid Mech. 147 (1984)
  1--38.
\newblock \href {https://doi.org/10.1017/S0022112084001968}
  {\path{doi:10.1017/S0022112084001968}}.

\bibitem{Armbruster1989}
D.~Armbruster, J.~Guckenheimer, P.~Holmes, Kuramoto--{S}ivashinsky dynamics on
  the center--unstable manifold, SIAM J. Appl. Math. 49~(3) (1989) 676--691.
\newblock \href {https://doi.org/10.1137/0149039} {\path{doi:10.1137/0149039}}.

\bibitem{Touze2011}
C.~Touz{\'e}, O.~Thomas, M.~Amabili, Transition to chaotic vibrations for
  harmonically forced perfect and imperfect circular plates, Int. J. Non-Linear
  Mech. 46~(1) (2011) 234--246.
\newblock \href {https://doi.org/10.1016/j.ijnonlinmec.2010.09.004}
  {\path{doi:10.1016/j.ijnonlinmec.2010.09.004}}.

\bibitem{Thompson2002}
J.~M.~T. Thompson, H.~B. Stewart, Nonlinear Dynamics and Chaos, 2nd Edition,
  Wiley, Chichester, 2002.

\bibitem{NayfehBalachandran1995}
A.~H. Nayfeh, B.~Balachandran, Applied Nonlinear Dynamics: Analytical,
  Computational, and Experimental Methods, Wiley, New York, 1995.

\bibitem{Moon1992}
F.~C. Moon, Chaotic and Fractal Dynamics: An Introduction for Applied
  Scientists and Engineers, Wiley, New York, 1992.

\bibitem{Strogatz2015}
S.~H. Strogatz, Nonlinear Dynamics and Chaos, 2nd Edition, Westview Press,
  Boulder, 2015.

\bibitem{Guckenheimer1983}
J.~Guckenheimer, P.~Holmes, Nonlinear Oscillations, Dynamical Systems, and
  Bifurcations of Vector Fields, Springer, New York, 1983.

\bibitem{Wiggins2003}
S.~Wiggins, Introduction to Applied Nonlinear Dynamical Systems and Chaos, 2nd
  Edition, Springer, New York, 2003.

\bibitem{Glendinning1994}
P.~Glendinning, Stability, Instability and Chaos: An Introduction to the Theory
  of Nonlinear Differential Equations, Cambridge University Press, Cambridge,
  1994.

\bibitem{Kuznetsov2004}
Y.~A. Kuznetsov, Elements of Applied Bifurcation Theory, 3rd Edition, Springer,
  New York, 2004.
\newblock \href {https://doi.org/10.1007/978-1-4757-3978-7}
  {\path{doi:10.1007/978-1-4757-3978-7}}.

\bibitem{Shilnikov1965}
L.~P. Shil'nikov, A case of the existence of a denumerable set of periodic
  motions, Sov. Math. Dokl. 6 (1965) 163--166.

\bibitem{Glendinning1984}
P.~Glendinning, C.~Sparrow, Local and global behavior near homoclinic orbits,
  J. Stat. Phys. 35~(5--6) (1984) 645--696.
\newblock \href {https://doi.org/10.1007/BF01010828}
  {\path{doi:10.1007/BF01010828}}.

\bibitem{Gaspard1984}
P.~Gaspard, R.~Kapral, G.~Nicolis, Bifurcation phenomena near homoclinic
  systems: A two-parameter analysis, J. Stat. Phys. 35~(5--6) (1984) 697--727.
\newblock \href {https://doi.org/10.1007/BF01010829}
  {\path{doi:10.1007/BF01010829}}.

\bibitem{Grebogi1983}
C.~Grebogi, E.~Ott, J.~A. Yorke, Crises, sudden changes in chaotic attractors,
  and transient chaos, Physica D 7~(1--3) (1983) 181--200.
\newblock \href {https://doi.org/10.1016/0167-2789(83)90126-4}
  {\path{doi:10.1016/0167-2789(83)90126-4}}.

\bibitem{NayfehMook1979}
A.~H. Nayfeh, D.~T. Mook, Nonlinear Oscillations, Wiley, New York, 1979.

\bibitem{Sanders2007}
J.~A. Sanders, F.~Verhulst, J.~Murdock, Averaging Methods in Nonlinear
  Dynamical Systems, 2nd Edition, Springer, New York, 2007.

\bibitem{Elliott1980}
J.~A. Elliott, Intrinsic nonlinear effects in vibrating strings, Am. J. Phys.
  48~(6) (1980) 478--480.
\newblock \href {https://doi.org/10.1119/1.11991} {\path{doi:10.1119/1.11991}}.

\bibitem{Tufillaro1992}
N.~B. Tufillaro, T.~Abbott, J.~Reilly, An Experimental Approach to Nonlinear
  Dynamics and Chaos, Addison-Wesley, Redwood City, CA, 1992.

\bibitem{Tufillaro2026data}
N.~B. Tufillaro, Three models of the whirling string through its route to
  chaos: run data and {MATLAB} code, Zenodo, [dataset] (2026).
\newblock \href {https://doi.org/10.5281/zenodo.23931253}
  {\path{doi:10.5281/zenodo.23931253}}.

\bibitem{Yang2016}
X.~Yang, Linear, first and second-order, unconditionally energy stable
  numerical schemes for the phase field model of homopolymer blends, J. Comput.
  Phys. 327 (2016) 294--316.

\bibitem{Shen2018}
J.~Shen, J.~Xu, J.~Yang, The scalar auxiliary variable ({SAV}) approach for
  gradient flows, J. Comput. Phys. 353 (2018) 407--416.

\bibitem{Henon1982}
M.~H{\'e}non, On the numerical computation of {P}oincar{\'e} maps, Physica D
  5~(2--3) (1982) 412--414.
\newblock \href {https://doi.org/10.1016/0167-2789(82)90034-3}
  {\path{doi:10.1016/0167-2789(82)90034-3}}.

\bibitem{Landau1937}
L.~D. Landau, On the theory of phase transitions, Zh. Eksp. Teor. Fiz. 7 (1937)
  19--32, {E}nglish translation in D. ter Haar (Ed.), Collected Papers of L. D.
  Landau, Pergamon, Oxford, 1965, pp.~193--216.

\bibitem{Neishtadt1984}
A.~I. Neishtadt, The separation of motions in systems with rapidly rotating
  phase, J. Appl. Math. Mech. 48~(2) (1984) 133--139.
\newblock \href {https://doi.org/10.1016/0021-8928(84)90078-9}
  {\path{doi:10.1016/0021-8928(84)90078-9}}.

\bibitem{Holmes1988}
P.~Holmes, J.~Marsden, J.~Scheurle, Exponentially small splittings of
  separatrices with applications to {KAM} theory and degenerate bifurcations,
  in: Hamiltonian Dynamical Systems, Vol.~81 of Contemporary Mathematics,
  American Mathematical Society, Providence, 1988, pp. 213--244.
\newblock \href {https://doi.org/10.1090/conm/081/986267}
  {\path{doi:10.1090/conm/081/986267}}.

\bibitem{Benettin1980}
G.~Benettin, L.~Galgani, A.~Giorgilli, J.-M. Strelcyn, Lyapunov characteristic
  exponents for smooth dynamical systems and for {H}amiltonian systems; a
  method for computing all of them. {P}art 1: Theory, Meccanica 15~(1) (1980)
  9--20.
\newblock \href {https://doi.org/10.1007/BF02128236}
  {\path{doi:10.1007/BF02128236}}.

\bibitem{Haller2016}
G.~Haller, S.~Ponsioen, Nonlinear normal modes and spectral submanifolds:
  existence, uniqueness and use in model reduction, Nonlinear Dyn. 86~(3)
  (2016) 1493--1534.
\newblock \href {https://doi.org/10.1007/s11071-016-2974-z}
  {\path{doi:10.1007/s11071-016-2974-z}}.

\bibitem{Breunung2018}
T.~Breunung, G.~Haller, Explicit backbone curves from spectral submanifolds of
  forced-damped nonlinear mechanical systems, Proc. R. Soc. A 474~(2213) (2018)
  20180083.
\newblock \href {https://doi.org/10.1098/rspa.2018.0083}
  {\path{doi:10.1098/rspa.2018.0083}}.

\bibitem{Li2022a}
M.~Li, S.~Jain, G.~Haller, Nonlinear analysis of forced mechanical systems with
  internal resonance using spectral submanifolds, {P}art {I}: Periodic response
  and forced response curve, Nonlinear Dyn. 110~(2) (2022) 1005--1043.
\newblock \href {https://doi.org/10.1007/s11071-022-07714-x}
  {\path{doi:10.1007/s11071-022-07714-x}}.

\bibitem{Li2022b}
M.~Li, G.~Haller, Nonlinear analysis of forced mechanical systems with internal
  resonance using spectral submanifolds, {P}art {II}: Bifurcation and
  quasi-periodic response, Nonlinear Dyn. 110~(2) (2022) 1045--1080.
\newblock \href {https://doi.org/10.1007/s11071-022-07476-6}
  {\path{doi:10.1007/s11071-022-07476-6}}.

\bibitem{Tufillaro1995}
N.~B. Tufillaro, P.~Wyckoff, R.~Brown, T.~Schreiber, T.~Molteno, Topological
  time-series analysis of a string experiment and its synchronized model, Phys.
  Rev. E 51~(1) (1995) 164--174.
\newblock \href {https://doi.org/10.1103/PhysRevE.51.164}
  {\path{doi:10.1103/PhysRevE.51.164}}.

\bibitem{Narducci1988}
L.~M. Narducci, N.~B. Abraham, Laser Physics and Laser Instabilities, World
  Scientific, Singapore, 1988.

\end{thebibliography}

\end{document}